\documentclass[fleqn,usenatbib]{mnras}

\usepackage[T1]{fontenc}

\DeclareRobustCommand{\VAN}[3]{#2}
\let\VANthebibliography\thebibliography
\def\thebibliography{\DeclareRobustCommand{\VAN}[3]{##3}\VANthebibliography}

\usepackage{graphicx}	
\usepackage{amsmath}	
\usepackage{amssymb}	
\usepackage{xcolor}	 
\usepackage{subcaption}

\newcommand{\afe}{$[\alpha/\rm{Fe}]$}
\newcommand{\feh}{$[\rm{Fe}/\rm{H}]$}
\newcommand{\msun}{M$_\odot$}
\newcommand{\bpass}{{\sc{bpass}}}

\title[Uncertainties in Population Synthesis: Spectra]{On the impact of spectral template uncertainties in synthetic stellar populations}

\author[Byrne et al.]{
C. M. Byrne\thanks{E-mail: conor.byrne@warwick.ac.uk (CMB)}
and E. R. Stanway
\\
Department of Physics, University of Warwick, Gibbet Hill Road, Coventry, CV4 7AL, UK\\
}

\date{Accepted XXX. Received YYY; in original form ZZZ}

\pubyear{2023}

\begin{document}
\label{firstpage}
\pagerange{\pageref{firstpage}-\pageref{lastpage}}
\maketitle

\begin{abstract}
Uncertainties in stellar population models, both in terms of stellar evolution and stellar spectra, translate into uncertainties in our interpretation of stellar populations in galaxies, since stars are the source of most of the light we receive from them. Observations by {\textit{JWST}} are revealing high-redshift galaxies in great detail, which must then be compared to models. One significant source of uncertainty is in the stellar spectra used to generate composite spectra of stellar populations, which are then compared to data. Confidence in theoretical models is important to enable reliable determination of the properties of these galaxies such as their ages and star formation history. Here we present a comparison of spectral synthesis carried out with 6 different stellar spectral libraries using the Binary Population and Spectral Synthesis (\bpass) framework. In photometric colours, the differences between theoretical libraries are relatively small ($<0.10$\,mag), similar to typical observational uncertainties on individual galaxy observations. Differences become more pronounced when detailed spectroscopic properties are examined. Predictions for spectral line indices can vary significantly, with equivalent widths differing by a factor of two in some cases. With these index strengths, some of the libraries yield predictions of ages and metallicities which are unphysical. Many spectral libraries lack wavelength coverage in the ultraviolet, which is of growing importance in the era of {\textit{JWST}} observations of distant galaxies, whose flux is dominated by hot, young stars.
\end{abstract}




\section{Introduction}

Fitting observed galaxy spectra, or less detailed spectral energy distributions (SEDs), to stellar population synthesis (SPS) models is a commonly used technique for determining the age, composition and star formation history of the stellar population in a distant galaxy where the individual stars cannot be resolved \citep[for a review of the SED fitting technique, see][]{Conroy13}. However, this method is sensitive to the choice of SED fitting code and underlying theoretical models, meaning that there is an intrinsic level of uncertainty associated with the results \citep[e.g.][]{2022arXiv221201915P}. As a consequence, these inferred properties are sensitive to the uncertainties in the population synthesis models, including the nature of the stellar population (single vs multiple stars), the stellar evolution models used (uncertainties in massive star evolution, convection, etc.) and the choice of stellar spectra (wavelength coverage, spectral resolution, atomic line lists, choice of radiative transfer code, etc.). Understanding how these uncertainties propagate through stellar population synthesis models and the degeneracies they introduce in inferred galaxy properties has implications for our understanding of galaxy evolution.



The use of SPS models to interpret the appearance of galaxies is an area of research which was pioneered by \citet{Tinsley68}. SPS models start with a set of stellar evolution models, weighted according to an initial mass function (IMF) which describes the proportion of stars with a given set of initial conditions that are expected to form in the stellar population. The evolution of these stars is then followed simultaneously to trace the evolution of the stellar population. Stellar spectra which match the atmospheric parameters of the stellar evolution models are combined in accordance with their relative contribution to the total luminosity of the stellar population at a given age and composition to produce a composite spectrum. Composite spectra for galaxies with complex star formation histories can be produced by combining simple stellar populations of different ages. Many different population and spectral synthesis models exist in the literature, each using a different combination of stellar evolution models and stellar spectral template libraries \citep[e.g.][]{BruzualCharlot03,2017PASA...34...58E,Maraston20}.

In young stellar populations, such as those in the distant Universe, massive stars are the dominant contributors to the spectrum. Massive stars are expected to be found often in multiple star systems \citep{2012Sci...337..444S}. Stellar multiplicity fractions also show an anti-correlation with stellar metallicity \citep[e.g.][]{ElBadry19,Moe19,Donada23}. Binary stellar interactions can significantly alter the composite spectrum of a stellar population \citep{Eldridge22Review}. For example, mass transfer can lead to rejuvenation of some stars, whereby accreting hydrogen-rich material spins up the star, mixing fresh hydrogen into the core. This allows them to stay on the Main Sequence longer, leading to a population appearing bluer than an equivalent population comprised solely of single stars of the same age. Therefore it is important to use SPS models which include binary effects when examining young galaxies.

There are multiple approaches to producing stellar spectra for population synthesis models. In some cases, empirical spectra are used, whereby spectra of observed stars are used to calculate the composite spectrum of the population \citep[e.g.][]{SanchezBlazquez06}. Using empirical spectra has the advantage of using real observations, and thus a spectrum of a stellar population is produced using a combination of stars as they actually appear in the sky. The limitation of this method is that it relies on detailed spectroscopic observations of bright, nearby stars and cannot necessarily be used to understand populations which have compositions or temperatures which are considerably different from stars close to the Sun. A related option is the use of semi-empirical spectra \citep[e.g.][]{Knowles21}, which modify the empirical spectra by applying corrections based on the changes in flux between slightly varied theoretical spectra. This expands the region of parameter space in which spectra based on empirical data can be used, but is still limited to conditions which are at least somewhat similar to the observed stars.

Another option is to solely use theoretical spectral libraries. These spectra are produced by determining the flux emitted from a model stellar atmosphere using a radiative transfer calculation, and can in principle be constructed for any arbitrary composition and effective temperature \citep[e.g.][]{Kurucz70}. This is more advantageous than empirical libraries for studying young stellar populations at high redshift, where the typical composition, mass and temperature of the stars in the population differ considerably to those commonly seen in the local environment and used in empirical libraries. A limitation of theoretical libraries is that they rely on tabulated atomic physics data (opacities and atomic absorption line strengths), some of which are not well constrained. Different software packages follow a variety of approaches to model atmosphere construction and radiative transfer calculation, and use specific linelists of atomic and molecular transitions, and thus can produce varied predictions for a theoretical spectrum, despite having the same grounding in atomic physics. 

An evaluation of the differences between theoretical and empirical spectral libraries in population synthesis has recently been carried out by \citet{Coelho20}. They find that photometric colours are sensitive to the parameter space coverage of the spectral grid, particularly for bluer filters. This is an unsurprising result given that the high-temperature portion of parameter space is poorly covered by the empirical spectra. One key derived quantity which differs between empirical and theoretical spectra is the metallicity determination of a stellar population. Theoretical libraries systematically infer a lower metallicity than those inferred by fitting with empirical libraries.

The abundance of chemical elements increases at different rates, as a consequence of the many processes underlying their formation \citep[e.g.][]{Kobayashi20}. Large quantities of $\alpha$-process elements such as oxygen and magnesium are synthesised in core collapse supernovae, an end state of massive star evolution, which happens on a comparatively short timescale. Meanwhile, Type Ia supernovae, which are an end state of low mass stellar evolution produce the majority of iron group elements (e.g. iron and nickel) on a much longer timescale. As a result, pristine, primordial gas will first become enriched in $\alpha$-process elements before there is significant growth in the abundance of iron group elements. This process, referred to as $\alpha$-enhancement, has been observed in deep spectroscopy of distant populations of galaxies at $z\sim$ 2 \citep{2016ApJ...826..159S,2022ApJ...925..116S} and $z\sim$ 3.4 \citep{2021MNRAS.505..903C}.

This non-uniform chemical evolution changes the behaviour and appearance of the stars which form in such an environment, adding another source of uncertainty to the interpretation of stellar populations. Young stellar populations in the early Universe, which will be observed in great numbers by {\textit{JWST}}, are one of the environments where this effect will be most pronounced. As a result, it is important that efforts to produce theoretical composite spectra of stellar populations formed in such an environment take into account the difference in the chemical abundance pattern, rather than simply scaling the patterns seen in nearby, high-metallicity environments.

Studies have been done on the effects that $\alpha$-enhanced compositions have on low-mass stellar evolution models, and the implications this has for stellar population synthesis. For example, \citet{Pietrinferni21} calculated a grid of $\alpha$-enhanced stellar evolution models using the BaSTI stellar evolution code \citep{Pietrinferni06,Vandenberg14}. That study set out to explore the impact of old, low-metallicity stellar populations in the Milky Way and the nearby environment, another regime where stars tend to be $\alpha$-enhanced. These are effectively the same stellar populations, just at vastly different ages. The young stellar populations in the distant Universe still contain the massive stars which are not present in old stellar populations, the evolution of which is significantly impacted by binary star interactions. In these works using the BaSTI code, the maximum initial stellar mass of the evolution models was 15\,\msun, and only single star evolution was considered. This limits the applicability of such models to young stellar populations in the distant Universe.

The effect of $\alpha$-enhancement on stellar spectra has also been explored in the context of stellar population synthesis. \citet{Vazdekis15} used $\alpha$-enhanced stellar evolution isochrones up to 10\,\msun\ at a resolution of $\sim2.5$\AA\ to measure the effect it has on intermediate and old stellar populations at optical wavelengths. A key finding was that $\alpha$-enhanced populations appear bluer at blue wavelengths than their Solar-scaled counterparts. \citet{Percival09} also investigated $\alpha$-enhanced stellar populations. They did so in the context of Galactic globular clusters, and found that such models could successfully reproduce the spread of ages and compositions seen in a large sample of Milky Way globular clusters. These studies neither accounted for the role of binary stellar evolution, nor explored the far-ultraviolet spectrum, both of which are key considerations for understanding young stellar populations. Stellar population synthesis models including binary stars and incorporating a single grid of $\alpha$-enhanced stellar spectra were calculated by \citet{Byrne22}. These models gave good agreement with the earlier studies, while also providing spectra which cover the UV spectrum. Several theoretical stellar spectral libraries have been produced which account for $\alpha$-enhancement. A number of these will be outlined in Section~\ref{sec:speclibs}. Each library uses differing methodology to generate its spectra, and thus the output spectra for a given set of stellar parameters will not be identical.

In general, each spectral library has been used in conjunction with a specific population synthesis model. To date, there has been no systematic study of different spectral libraries using a fixed binary stellar population synthesis model to evaluate how the choice of spectral library can change the predicted appearance of the same, fixed assortment of stars.

In this paper, we explore the impact that choices of spectral library can have on the observed properties of a stellar population, and the implications that they have for inferred properties of distant galaxies. Section~\ref{sec:popsynth} outlines the population synthesis methods used, while Section~\ref{sec:specsynth} outlines the various spectral template libraries considered and the methods used to normalise them for best comparison. Results are presented in Section~\ref{sec:output}, with further discussion in Section~\ref{sec:discuss}, future developments outlined in Section~\ref{sec:future} and concluding remarks in Section~\ref{sec:conclusions}.

\section{Population Synthesis}
\label{sec:popsynth}

Binary Populations and Spectral Synthesis \citep[\bpass,][]{2017PASA...34...58E,2018MNRAS.479...75S,Byrne22} is a framework of stellar population and spectral synthesis codes, notably including binary stellar evolution and the impact of binary interactions, tracing the evolution of simple stellar population from an age of 1\,Myr up to 100\,Gyr using a grid of detailed single and binary stellar evolution models. 

Individual stellar evolution models are combined using a \cite{2001MNRAS.322..231K} initial mass function extending from 0.1 to 300 M$_\odot$ to produce a stellar population of mass $10^6$\,M$_\odot$. Empirical estimates from \cite{2017ApJS..230...15M} of the period distribution, mass ratio distribution and mass-dependent binary fraction are used to weight the stellar models. The stellar evolution models are then matched according to their temperature, surface gravity and composition to an appropriate stellar atmosphere model to produce an integrated spectrum for the stellar population. This is usually achieved by selecting the appropriate template library based on the type of star (Main Sequence/Giant Branch, White Dwarf, Wolf-Rayet star, etc.) and interpolating  between spectra in the library which bracket the stellar evolution model in effective temperature and surface gravity. Models which lie slightly outside the parameter coverage of the spectral template library are assigned the spectrum of the nearest template in the spectrum, rather than using extrapolation. The stellar evolution models assume a Solar abundance of \citet{Grevesse93} with a metal mass fraction Z$_\odot=0.02$ and a hydrogen mass fraction scaled by $X = 0.75-2.5Z$. The detailed stellar models trace the radial distribution and nucleosynthesis of H, He, C, N, O, Ne, Mg, Si and Fe. Detailed descriptions of the stellar evolution physics and construction of the models can be found in \citet{2017PASA...34...58E}.

\bpass\ has been used extensively to predict and interpret the appearance of massive star populations \citep[e.g.][]{2021MNRAS.505.2447P,2021MNRAS.505..903C,2021MNRAS.501.3289V,2021ApJ...908...87N,2020MNRAS.499.3819M,2020MNRAS.498.1347S,2020ApJ...900..118N,2020ApJ...899..117S,2020MNRAS.495.4430T,2020MNRAS.495.1501C,2020ApJ...896..164D,2020MNRAS.493.6079W,2020MNRAS.491.3479C,2020ApJ...888L..11S} as well as a number of studies of low-mass stellar populations \citep[e.g.][]{2017PASA...34...58E,2018MNRAS.479...75S,Byrne21}. The inclusion of binary stellar evolution pathways makes it a powerful tool for interpreting stellar populations of all kinds, but particularly those seen in distant galaxies. By keeping the properties of the stellar population fixed and changing the stellar spectral template library, it is possible to systematically evaluate a selection of  spectral template libraries and the differences they make to the composite spectrum of a given stellar population.

\section{Spectral Synthesis}
\label{sec:specsynth}
\subsection{Stellar Spectral Libraries}
\label{sec:speclibs}

Six spectral libraries were selected for evaluation within \bpass. These spectral libraries were chosen based on a number of criteria, such as being publicly available, having broad parameter space coverage, having spectra with non-Solar \afe, and use in current or previous versions of \bpass\ and other population synthesis codes. A brief description of each of these libraries and their properties are provided below, with important characteristics shown in Table~\ref{tab:libraries}. In each case, the surface gravity range covered refers to the widest possible range. In reality the low gravity limit is not fixed, and increases with increasing effective temperature.

\begin{table*}
    \centering
    \caption{Key properties of theoretical stellar spectral libraries which include $\alpha$-enhancement and/or have been used in \bpass.}
    \label{tab:libraries}
    \begin{tabular}{c|c|c|c|c}
    \hline
    Name   & Wavelength [\AA]  & T$_{\rm{eff}}$ [K] & Gravity [cgs] & \feh\ values (step)\\ 
    Reference & Resolution & Solar Composition &  & \afe\ values (step) \\\hline

    AP      & 2000-25000     & $3500\le$ T$_{\rm{eff}}\le30000^*$ & $0\le\log(g)\le5$ & -5.0 -- 0.5 (0.25)\\ \citet{AllendePrieto18Paper}& R $\simeq30000$&\citet{Asplund05}& & -1.0 -- 1.0 (0.25) \\ \hline

    BaSeL      & $\sim$1785-100000    & $2000\le$ T$_{\rm{eff}}\le50000^*$ &$-1.02\le\log(g)\le5.5$ & -2.0 -- 0.5 (0.5)\\ \citet{Westera02}& R $\simeq110$ (optical)& semi-empirical & & 0.0 only \\ \hline

    C3K      & 100-20000    & $2500\le$ T$_{\rm{eff}}\le50000^*$ &$-1\le\log(g)\le5.5$& -4.0 -- 0.5 (0.25) \\ \citet{Choi16}& R $=10000$&\citet{Asplund09}& & -0.2 -- 0.6 (0.2) \\ \hline

    CKC14      & 100-100000    & $2500\le$ T$_{\rm{eff}}\le25000$ &$-1.02\le\log(g)\le5.5$ & -2.5 -- 0.5 (0.25) \\ \citet{Conroy14}& R $=3000$ (optical)&\citet{Asplund09}& & 0.0 only\\ \hline

    Coelho      & $\sim$2500-9000    & $3000\le$ T$_{\rm{eff}}\le20000$ &$-0.5\le\log(g)\le5.5$ & -1.3 -- 0.2 (variable)\\ \citet{Coelho14} & R $=20000$&\citet{Grevesse98}& &\ 0.0 and +0.4  \\ \hline

    sMILES theoretical models      & $\sim$1680-9000    & $3500\le$ T$_{\rm{eff}}\le10000$ &$0\le\log(g)\le5$ & -2.5 -- 0.5 (0.5) \\ \citet{Knowles21} & $\delta\lambda=0.05$\AA&\citet{Asplund05}& &-0.25 -- 0.75 (0.25) \\ \hline
    
    \multicolumn{5}{l}{$^*$BPASS uses the Main Sequence spectral libraries up to 25000\,K before switching to an OB star grid at higher effective temperatures.}\\
    
    \end{tabular}
\end{table*}

\subsubsection{AP - Allende Prieto et al. (2018)}

The \citet{AllendePrieto18Paper} (AP) spectral library covers a broad range of composition parameter space, with spectra covering temperatures from 3\,500\,K to 30\,000\,K, \feh\ spanning 5.5 dex from -5.0 to 0.5 in 0.25 dex increments and a 2 dex variation in \afe\ from -1 to +1, also in increments of 0.25 dex. The wavelength coverage extends from the near ultraviolet (2000\,\AA) to the near infrared (2.5\,$\mu$m). The Solar composition of \citet{Asplund05} is used as the base elemental mixture.

These spectra were computed using the {\sc{atlas9}} \citep{Kurucz92} model atmospheres of \citet{Meszaros12}. Spectral synthesis was carried out with the {\sc{ass}}$\epsilon${\sc{t}} radiative transfer code \citep[][]{Koesterke08,Koesterke09}.

The wide parameter space coverage of this spectral library makes it a useful direct comparator to the existing CKC14 and C3K libraries used in \bpass\ (see below), while the fine increments of \afe\ also make it convenient for exploring the effects of $\alpha$-enhancement in the context of population synthesis.

It should be noted that while some spectra in this grid cover temperatures between 25\,000\,K and 30\,000\,K, \bpass\ presently uses a grid of OB stellar spectra calculated using {\sc{WM-BASIC}} \citep{PauldrachWMBASIC} for main sequence stars hotter than 25\,000\,K, so the hot spectra in the AP library are not used in the spectral synthesis in this work.

\subsubsection{BaSeL -  Westera et al. (2002)}

The BaSeL spectral libraries \citep[][]{Lejeune98,Westera02} have been a popular choice in population synthesis codes. For example, the widely used \citet{BruzualCharlot03} single-star population synthesis models make use of the BaSeL spectra. The BaSeL spectra were also used in \bpass\ up to and including v2.1 \citep{2017PASA...34...58E}. While these spectra cover a broad region of parameter space, the spectral resolution is rather poor in comparison to the outputs of more recent stellar atmosphere models. BaSeL spectra were only used up to an effective temperature of 25\,000\,K in previous versions of \bpass\, so this is maintained here.

\subsubsection{CKC14 - Conroy et al.(2014)}

The CKC14 spectral library \citep[][]{Conroy14} was used as the Main Sequence and Giant Branch spectral library in \bpass\ v2.2 \citet{2018MNRAS.479...75S}. These stellar atmospheres and spectra were constructed using the {\sc{atlas12}} \citep{Kurucz05,Castelli05} and {\sc{synthe}} \citep{Kurucz93} codes respectively. The effective temperature and surface gravity coverage is comparable to the BaSeL library, but is likewise restricted to a uniform abundance scaling of the Solar mixture. The CKC14 spectra use the \citet{Asplund09} mixture as the base element distribution.

As well as \bpass, the CKC14 spectra have been implemented in the {\sc{fsps}} population and spectral synthesis code \citep{Conroy10} and the related tool {\texttt{prospector}} \citep{JohnsonPROSPECTOR}.

\subsubsection{C3K - Choi et al. (2016)}

The C3K spectra \citep[][]{Choi16} are a successor to the CKC14 spectra, with updated Kurucz linelists, once again using {\sc{atlas12}} and {\sc{synthe}}. The parameter space coverage and resolution are comparable to the CKC14 spectra, but cover a broader range of metallicities (\feh\ down to -4) and are notable for the inclusion of $\alpha$-enhanced spectra (\afe\ from $-0.2$ to $+0.6$ in 0.2 dex increments). As with the CKC14 library, the \citet{Asplund09} solar mixture is used as the base composition for this library. These spectra were introduced to \bpass\ in v2.3 \citep{Byrne22}. As with the AP library, the hottest stars in this library are not used in this work.

\subsubsection{Coelho - Coelho (2014)}

\cite{Coelho14} produced a small number of spectral libraries exploring the impact of $\alpha$-enhancement on the stellar spectra. Model atmospheres were produced using {\sc{atlas9}}, and spectral synthesis was carried out with {\sc{synthe}}. A lower temperature limit of 3000\,K was adopted by the author owing to the presence of dust in stars cooler than this and the lack of VO molecular lines in their line lists, which would make such spectra unreliable.

Spectra were generated at both \afe=0 and \afe=0.4, with the $\alpha$-enhanced spectra being produced at `fixed iron' and `fixed Z', whereby the $\alpha$-enhanced compositions are chosen such that they match \feh\ or $Z$ of the \afe=0 spectra respectively. In these libraries, the Solar abundance mixture of \citet{Grevesse98} is adopted as the base composition.

\subsubsection{sMILES theoretical models - Knowles et al. (2021)}

The {\sc{MILES}} \citep[\textbf{M}edium Resolution \textbf{I}saac Newton Telescope \textbf{L}ibrary of \textbf{E}mpirical \textbf{S}pectra,][]{SanchezBlazquez06} stellar spectral library consists of $\sim1000$ flux calibrated stars between 3500--7500\AA. The empirical spectra provide good coverage of effective temperature and surface gravity, but are naturally constrained in composition by the composition of the stars which were observed in the programme. In order to explore the effects of arbitrary abundance changes, the sMILES \citep[semi-analytical MILES,][]{Knowles21} study produced a number of theoretical spectral libraries to derive semi-analytical corrections to the observed spectra and generate pseudo-realistic observed spectra with a modified composition. These spectra were generated with the same underlying {\sc{atlas9}} stellar atmosphere models and {\sc{ass$\epsilon$t}} radiative transfer code as \citet{AllendePrieto18Paper}, using the \citet{Asplund05} Solar composition. A number of compositions are available, covering a wide range of values of \feh\ and \afe. Additionally, these theoretical sMILES spectra include libraries with variations in carbon abundance. These carbon-variable libraries have not been explored in this work.

The major limiting factor of the sMILES spectral library is that the maximum effective temperature is 10\,000\,K. This is sufficient for the studies of old, low-mass stellar populations, where most stars are cool. However, to study younger stellar populations, which generally contain hotter stars, a wider effective temperature coverage is needed. References to the sMILES library throughout the remainder of the paper will refer to the theoretical spectral templates generated for the sMILES study, and not the resultant semi-empirical stellar spectra.

\subsection{Normalisation}
\label{sec:standards}

These spectral libraries use varying definitions of Solar metallicity and composition mixtures, cover different wavelength ranges, and have differing coverage of temperature and surface gravity parameter space. It is important to find a method to normalise all libraries to an equivalent bolometric luminosity. In the first instance, the total metallicity mass fraction, $Z$, is calculated for each composition sub-grid in each spectral library. This is computed based on the adopted choice of Solar metallicity and the stated values of \feh\ and \afe. Interpolated libraries of spectra are then produced to match the values of $Z$ which are used in the stellar evolution models in \bpass. This ensure that the value of $Z$ in the stellar spectra matches the underlying stellar evolution models they are applied to. Given the different choices of Solar abundance profile by each grid, the relative proportions of individual elements will vary somewhat. All spectra were, where necessary, converted to vacuum wavelengths for consistency with the rest of \bpass.

The next decision is a choice of common normalisation so that each library will return a spectrum with the same bolometric flux for a given value of temperature, gravity and composition. The CKC14 and BaSeL spectra, with their large parameter space coverage and wavelength range were normalised to have a total flux of 1\,L$_{\odot}$, as is canonically the case in \bpass, and then multiplied by the luminosity of the star to get the total flux. Without an additional normalisation step, the other spectral libraries, would have a larger implied luminosity since 1\,L$_\odot$ of luminosity would be contained within a narrower wavelength range.

\begin{figure*}
    \centering
    \includegraphics[width=0.98\textwidth]{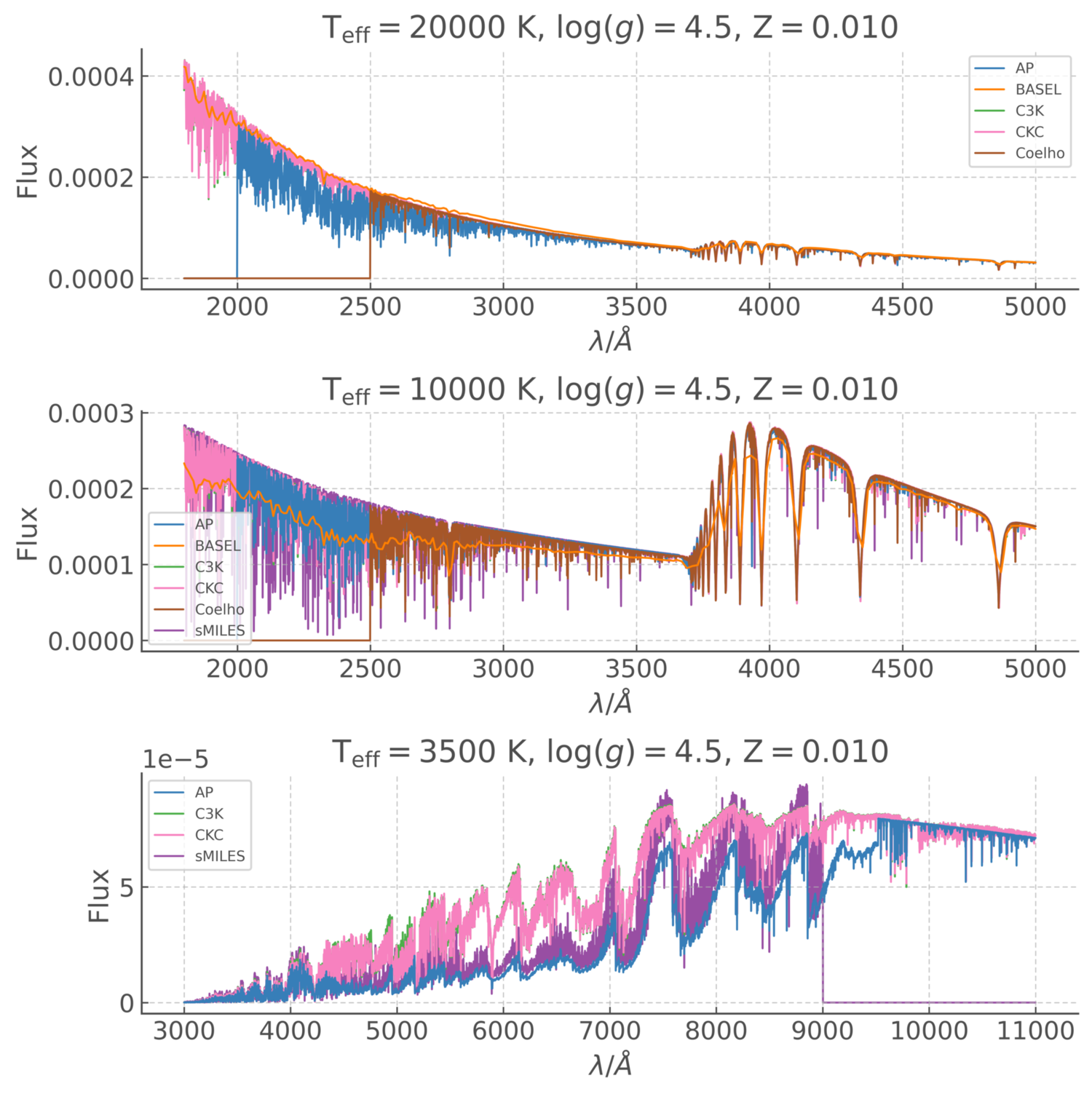}
    \caption{Portions of hot (20,000\,K, upper panel), intermediate (10,000\,K, middle panel) and cool (3,500\,K, lower panel) stellar spectra with a total metallicity mass fraction of $Z=0.01$ and a surface gravity of $\log(g/\mathrm{cm}\,\mathrm{s}^{-2})=4.5$ from each spectral library after the normalisation procedures have been carried out and the spectra have been resampled at a fixed 1\,\AA\ resolution. The two uppermost panels show the near-ultraviolet and optical spectrum, while the lower panel shows the optical and near-infrared.}
    \label{fig:stellar}
\end{figure*}

To account for this, the flux for a given spectra in the Johnson-Cousins V, I and J filters were computed, and the AP, C3K, Coelho and sMILES spectra were normalised using the following scheme.

\begin{enumerate}
    \item  All spectra with $T_{\mathrm{eff}}>6500$\,K are normalised to match the V-band flux of the equivalent CKC14 spectrum.
    \item  AP and C3K spectra with $T_{\mathrm{eff}}\le6500$\,K are normalised to match the J-band flux of the equivalent CKC14 spectrum.
    \item  C14 and sMILES spectra with $4500< T_{\mathrm{eff}}/{\mathrm{K}}\le6500$ are normalised to match the I-band flux of the equivalent CKC14 spectrum, since their wavelength coverage does not cover the J band.
    \item  C14 and sMILES spectra with $T_{\mathrm{eff}}\le4500$\,K are normalised to match flux of the equivalent CKC14 spectrum, in three specific wavelength bands (7450--7550\,\AA, 8075--8175\,\AA\ and 8750--8850\,\AA) owing to the strong molecular absorption bands in these cool star spectra and the challenge this presented in choosing a single filter profile for normalisation.
    \item  An interpolated CKC14 spectrum is computed to determine the normalisation required for spectra in the other libraries which have $T_{\mathrm{eff}}$ and $\log(g)$ combinations not already present in the CKC14 library.
\end{enumerate}

This ensures that the integrated spectra sum to a value of less than 1, owing to the shorter wavelength coverage, but that the bolometric flux should be comparable between different libraries, allowing the best comparison of outputs. Fig~\ref{fig:stellar} shows some example $Z=0.01$, \afe=0 stellar spectra at a high (20,000\,K), moderate (10,000\,K) and low (3,500\,K) effective temperature in the upper, middle and lower panels respectively. The two warmer spectra show the near ultraviolet and blue optical portion, while the cool stellar spectrum shows the red and near infrared portion of the spectrum. These fragments of spectra illustrate the overall success in normalisation of the spectra to match their fluxes in the wavelength ranges that are common to each spectral library, while also highlighting the issues and challenges in getting an exact match. In most cases, the underlying continuum flux is in agreement, but there can be significant differences in the strength of specific spectral line features. 

The low spectral resolution of the BaSeL spectra is evident in the upper two panels, with most spectral features being heavily smoothed. The short wavelength limits of the AP and Coelho libraries are also evident in these two panels, while the absence of the sMILES library in the top panel is indicative of its narrow effective temperature coverage. The lower panel highlights the difficulties in normalising and comparing cool star spectra. Different predictions for the strength of the molecular absorption bands lead to considerably different fluxes. Additionally, there is a very sudden break seen in the AP spectrum at $\sim$9500\AA, which is not seen in the other spectral libraries at this temperature. This break gets weaker with increasing effective temperature and is barely noticeable above an effective temperature of 4,000\,K. This does pose a challenge in producing an effective bolometric normalisation. Normalising to the flux in the Johnson-Cousins J filter, which was the option chosen for this work, gives a good match to the infrared continuum of the CKC14 and C3K spectra, but leads to a lower predicted flux in the optical than the other libraries. Alternatively, normalising to match the flux maxima between the molecular absorption bands (as was done for the sMILES and Coelho cool star spectra due to the lack of coverage of the J filter) would better match the optical flux, but would significantly overpredict the infrared flux. The physical origin of this spectral break is unclear. There will be a noticeable divergence between spectral synthesis results in stellar populations where the flux is dominated by cool stars, either a small number of very luminous red supergiants at young stellar ages, or a large number of cool, low-mass Main Sequence stars at late ages.

The same bolometric normalisation procedure was applied to the $\alpha$-enhanced libraries, using the \afe=0 CKC14 as the reference spectra. While this is not an ideal match, in the absence of full (1-100\,000\AA) $\alpha$-enhanced spectra, this was deemed the most appropriate choice of normalisation, as the overall flux in the covered wavelength ranges is similar to the \afe\ case for the same stellar parameters.

\subsection{Combining Population and Spectral Synthesis}

Having constructed standardised libraries of stellar spectra, each one was used to produce \bpass\ spectral synthesis outputs, for each of the \bpass\ metallicities which are within the range of the metallicity of the original spectral libraries. \bpass\ spectral synthesis outputs were also calculated for an $\alpha$-enhancement of 0.4 dex for the spectral libraries where this was a possibility, namely the AP, C3K, Coelho and sMILES libraries. Aside from the varying spectral library, each calculation uses the fixed fiducial set of \bpass\ input parameters ($10^6$\,M$_\odot$ initial population mass, \citet{2001MNRAS.322..231K} IMF from 0.1 to 300\,M$_\odot$ and initial binary parameter distributions from \citet{2017ApJS..230...15M}). A standard set of \bpass\ outputs is generated, one for each metallicity, spectral library choice and \afe\ value, giving a total of 116 variants.

\section{Model Outputs}
\label{sec:output}
\subsection{Photometric Colours}

\begin{figure*}
    \centering
    \includegraphics[width=0.96\textwidth]{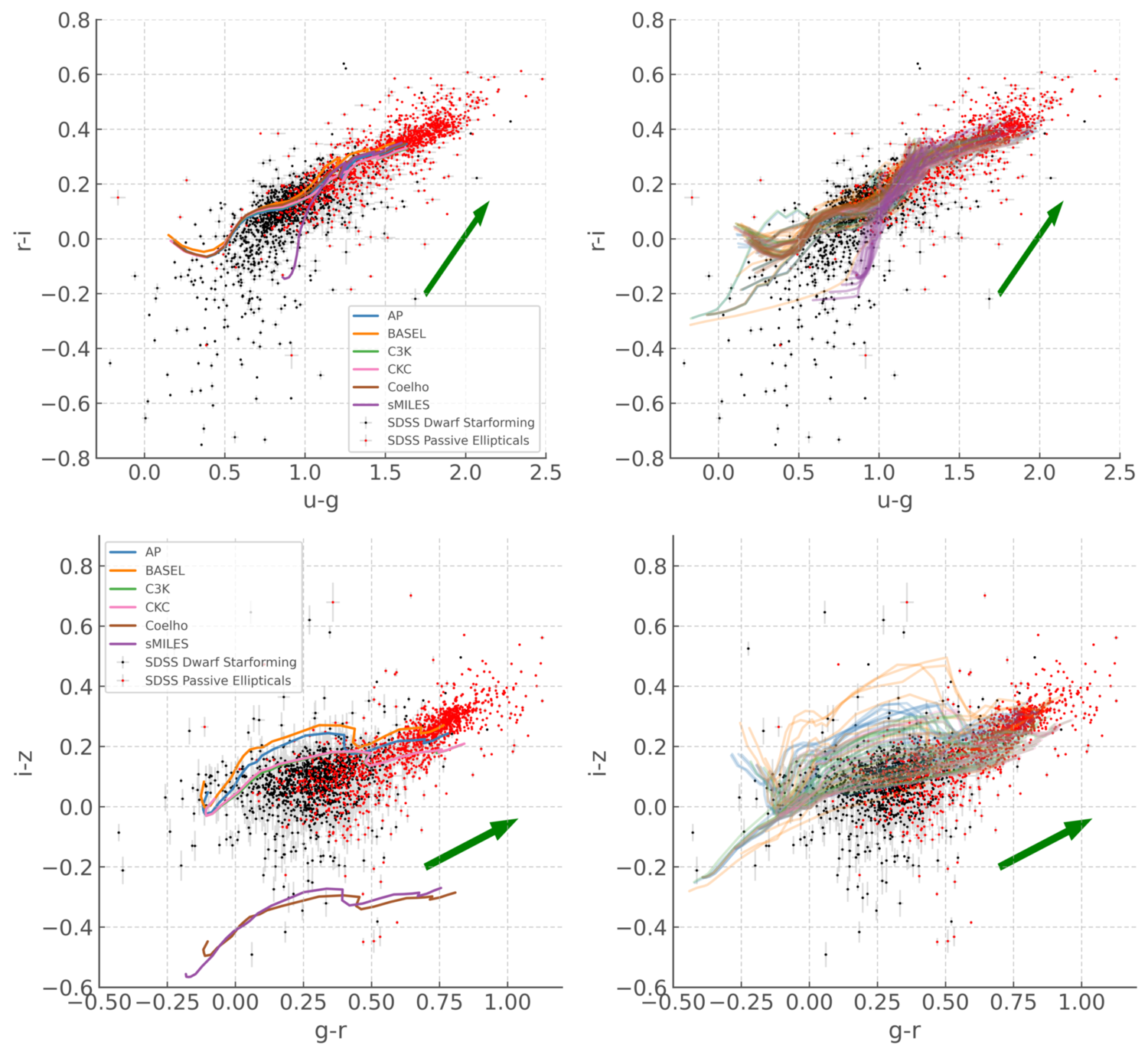}
    \caption{Colour-colour diagrams in $u-g$ vs $r-i$ (upper panels) and $g-r$ vs $i-z$ (lower panels). Photometry for nearby ($z<0.02$) dwarf star-forming (black points) and passive elliptical (red points) galaxies from SDSS are shown, along with their uncertainties.  \bpass\ spectral synthesis results for $Z=0.008$, \afe=0.0 for each library are shown as a function of population age in the left-hand panels, while the right-hand panels plot the results for all combinations of metallicity, spectral library and \afe\ of 0.0 and 0.4. The tracks associated with the Coelho and sMILES libraries are not plotted in the lower right panel due to their lack of infrared wavelength coverage giving spurious values for $i-z$, as shown in the lower left panel. An extinction vector ($A_{\mathrm{V}}=1.0$) is shown in green in each panel. The colours typically evolve from the lower left (young ages) to upper right (old ages) in the diagram.}
    \label{fig:colourcolour_sdss}
\end{figure*}

Photometric colours can be obtained for a very large number of objects simultaneously, with a relatively small cost in observing time compared to spectroscopy. Fig~\ref{fig:colourcolour_sdss} uses colour-colour diagrams to compare the spectral synthesis results using the different libraries to photometric observations of nearby ($z<0.02$) spectroscopically-confirmed galaxies identified by the Sloan Digital Sky Survey \citep[SDSS DR7,][]{SDSS_DR7} whose properties have been characterised by the MPA-JHU analysis. We restrict the sample to galaxies with photometric uncertainties less than 0.05\,mag. In particular, two observational populations were selected; dwarf star-forming galaxies ($\log(\mathrm{M}_\star/\mathrm{M}_\odot)<10^8$ and assigned  `{\sc{starforming}}' or `{\sc{starburst}}' spectroscopic subclasses in the SDSS data) and passive galaxies (galaxies not assigned a star-forming or AGN subclassification). This allows us to evaluate the models against galaxies which are dominated either by young or old stellar populations respectively.

The stellar mass and star formation rates of these galaxies were extracted from the MPA-JHU data tables on the SDSS database. As such, these values, which are derived through a combination of star formation rate indicators and spectral energy distribution fitting, have been calibrated using models other than \bpass. The stellar masses were determined by the MPA-JHU team using the Bayesian methodology and model grids described in \citet{Kauffmann03}, and they computed the star formation rates from nebular emission lines following the method described by \citet{Brinchmann04}.

These derived values will differ from the estimates that would be made had they been fitted with \bpass. As \citet{2022MNRAS.514.5706J}, \citet{2022arXiv221201915P} and others have shown, the stellar masses recovered through different SED fitting templates are highly robust, showing systematic offsets of $\Delta$log(M/M$_\odot$)$<$0.2, while star formation rate indicators can show a somewhat larger variation.

The left-hand panels of Fig~\ref{fig:colourcolour_sdss} compare the galaxy photometry to the colours of a simple stellar population with metallicity $Z=0.008$ as it ages from 20\,Myr to 16\,Gyr. The stellar population is bluer at young ages, so the time evolution of each line typically proceeds from the left-hand side to the right-hand side of the diagram. In the right-hand panels, the spectral synthesis results for all libraries and all metallicities (including $\alpha$-enhanced compositions) are plotted, to illustrate the full spread in colours predicted by the models. In the case of the $u-g$ vs $r-i$ colour-colour space, the colours predicted by \bpass\ cover the locus occupied by the observations very well. With the exception of the sMILES library at young ages (whose flux in $u$ will be underestimated due to the absence of hot star spectra), there is good agreement between the different libraries, with a spread of less than 0.1\,mag (0.08\,mag in $u-g$, 0.09\,mag in $r-i$), which is similar to typical uncertainties and spread in the observations. 

In the $g-r$ vs $i-z$ case, the long wavelength limit of 9000\,\AA\ in the sMILES and Coelho libraries leads to spurious values for $i-z$ since the $z$ filter extends across wavelengths up to $\sim10500$\,\AA, as can be seen clearly in the lower-left panel. These libraries are excluded from the lower right-hand panel for clarity. For the remaining 4 libraries, the spread in colour-colour space is slightly larger ($\sim$0.1\,mag, occasionally a little more) but is still generally consistent with observational scatter. Looking at both of the lower panels, it is clear that the theoretical results from \bpass\ consistently lie above  the observations in $i-z$. Extinction vectors ($A_{\mathrm{V}}=1.0$) are shown in each panel; invoking a modest amount of dust extinction leads to better agreement between the theoretical models and the observations.

A possible contributing factor to the offset in $i-z$ is the theoretical uncertainty in stellar evolution predictions for red supergiant stars \citep[e.g.][]{Beasor21,Davies21,Massey21}. As these stars are very luminous, uncertainties in their evolution could make a perceptible shift in the red and infrared flux of the stellar population.

In addition the bolometric normalisation of cool stellar spectra is challenging, given that their optical luminosities are strongly dependent on the assumed absorption species line-lists and line strengths. As was discussed in Section~\ref{sec:standards}, this makes comparative cross-normalisation of spectra in the optical extremely challenging, while many observational constraints and theoretical atmospheres for these stars do not extend significantly into the infrared. It should be noted that ground-based observations in the $i$ and $z$ filters are also challenging due to the poor quantum efficiency of CCD detectors and reliably accounting for atmospheric contamination.

\subsection{Mass-to-light ratios}

\begin{figure*}
    \centering
    \includegraphics[width=0.98\textwidth]{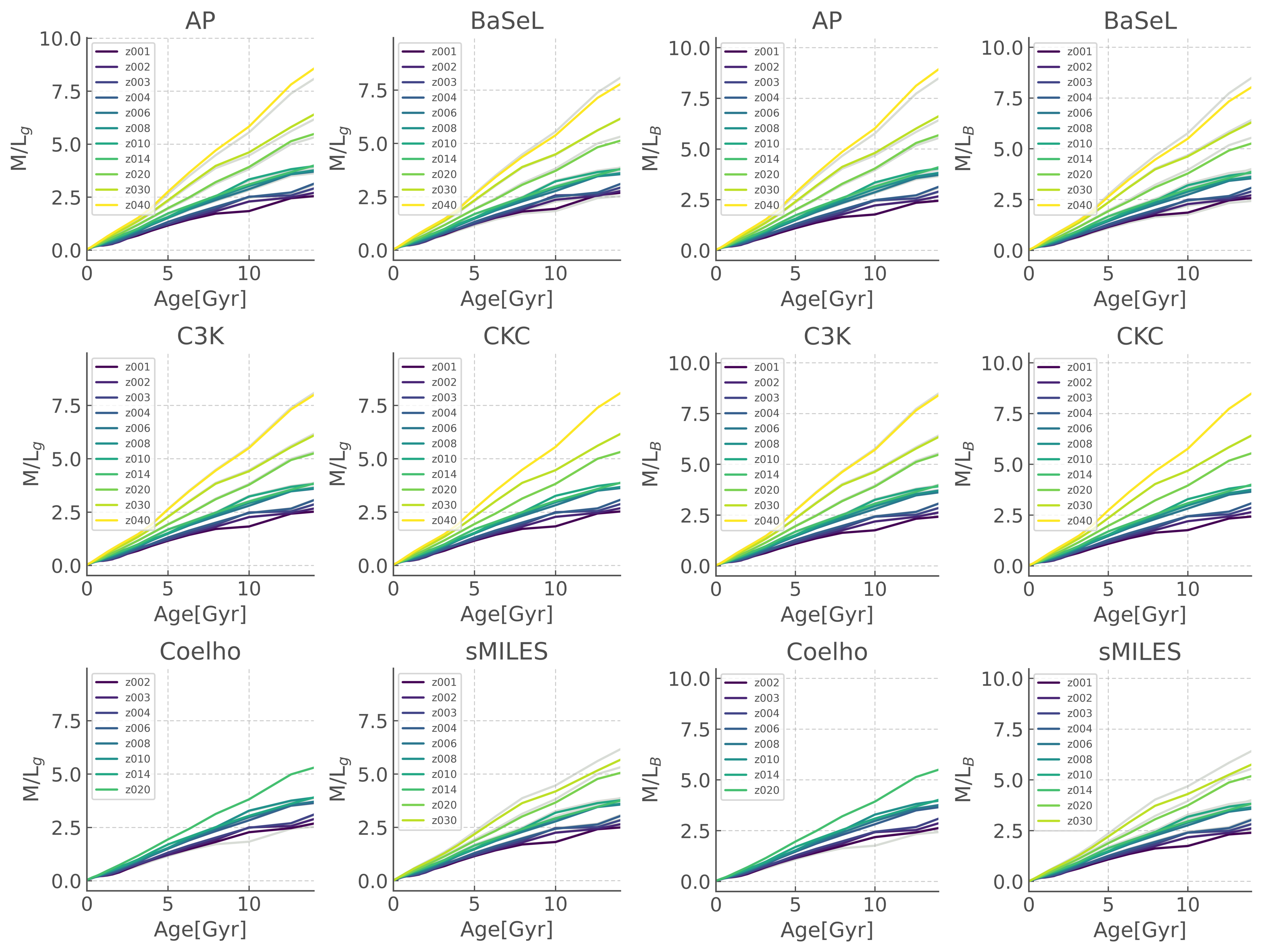}
    \caption{Mass to light ratios in the $g$ (left-hand columns) and B (right-hand columns) filters as a function of age, with each metallicity represented by a different colour and each spectral library in a separate panel. The lines corresponding to the CKC14 library are plotted in light grey in each panel for comparison.}
    \label{fig:MtoL}
\end{figure*}

The observed stellar mass-to-light ratio ($M/L$) can be a useful quantity, particularly for galaxy surveys in which photometric and spectroscopic coverage is insufficient to permit robust model fitting. Instead luminosity in a single photometric band can be used as a convenient proxy for stellar mass \citep[e.g.][]{2004Natur.430..181G}. Fig~\ref{fig:MtoL} shows the mass-to-light ratio in the SDSS $g$ and Johnson-Cousins $B$ band filters for stellar populations generated using each spectral library as a function of age and metallicity. Each panel also reproduces the lines corresponding to our reference CKC14 library in light grey. In general, the differences in M/L are very small ($<\pm2.5$ per cent from the reference library at ages of 1\,Gyr or more) at a given age and metallicity, indicating that this is a robust quantity generally independent of the choice of spectral library. 

For ages of 1\,Gyr or more, the C3K and Coelho libraries differ from the CKC14 reference library by no more than 1.25 per cent at any metallicity. At the highest metallicities under consideration ($Z=0.040$), the AP library predicts a slightly higher $M/L$ (up to 7 per cent in $g$, 6 per cent in $B$ at the highest age considered, 16\,Gyr) than the CKC14 library, with the offset increasing with age. By contrast the SMILES library predicts a lower $M/L$ at high metallicities and late ages (15 per cent lower in $g$ and 18 per cent in $B$ at 16\,Gyr). The same pattern is seen in both $g$- and $B$-band derived values. The BaSeL library-derived values show a smaller offset from CKC14 (the largest offsets are about 6 per cent in both bands, seen at late ages and/or low metallicities). 

\subsection{4000 Angstrom break}

\begin{figure}
    \centering
    \includegraphics[width=0.48\textwidth]{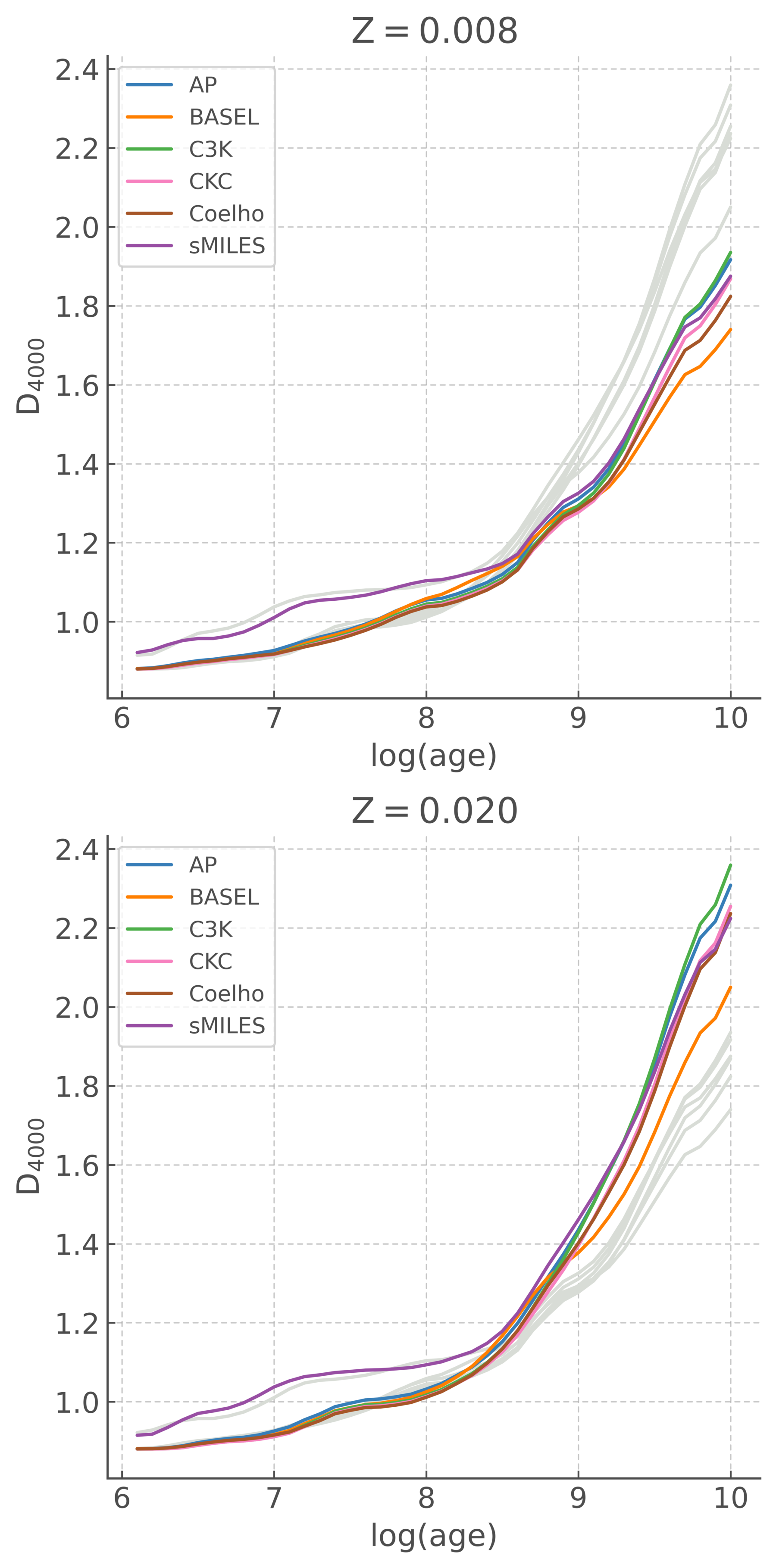}
    \caption{Predicted strength of the 4000\,\AA\ break as a function of stellar age for each spectral library at two different metallicities, $Z=0.008$ (coloured lines in upper panel, grey lines in lower panel) and $Z=0.020$ (coloured lines in lower panel, grey lines in upper panel).}
    \label{fig:4000break}
\end{figure}

The spectral break at 4000\,\AA\ is a powerful diagnostic of stellar population age. It captures the combined impact of hydrogen and metal absorption edges seen in the spectra of mid-sized stars. Thus it emerges as the dominance of more massive stars in young stellar populations wanes. We demonstrate the impact of assumed stellar library on this quantity in Fig~\ref{fig:4000break}, where we have calculated the break strength, $D_{4000}$ defined as the ratio of the flux in the spectrum between $4000-4100$\,\AA\ and $3850-3950$\,\AA. The differences between the stellar libraries are less than 2.5 per cent at ages less than 100\,Myr. The exception to this is the sMILES library, which is temperature-limited to 10,000\,K, and thus unreliable for young stellar populations. At ages later than 100\,Myr, when the population is dominated by cooler stars, spectra derived from the sMILES library are consistent with most of the other libraries. 
The BaSeL library consistently predicts the smallest break, while the AP and C3K libraries show the largest values for $D_{4000}$, with the other libraries lying in between these values. At an age of 10 Gyrs, the spread between BaSeL- and C3K-derived values is $\Delta D_{4000}=12$ per cent at $Z=0.030$, while all other libraries lie within $\Delta D_{4000}=4.7$ per cent of each other.

Since the difference between spectral libraries cannot arise from different temperatures or luminosities of stars in the population, it can only reflect a different treatment of the hydrogen, iron and other metal absorption line complexes that contribute to the break. In particular this is affected by the completeness and calibration of the metal species line lists incorporated in the spectral synthesis code used to generate each set of underlying stellar templates.

\subsection{Spectral indices as indicators of age and metallicity}
\label{sec:indices}
Spectral indices are pseudo-equivalent width line strength measurements that permit information on the stellar population to be extracted from low spectral resolution, or low signal-to-noise spectroscopy for which a full spectral fit is impossible. In each case, a pseudo-continuum region is defined and a ratio taken against the integrated flux of a region deemed to contain one or more absorption features. The spectral index list defined by the Lick Observatory has traditionally been used to classify and characterise nearby stars, and old simple stellar populations such as globular clusters \citep[][]{Faber85,Burstein86,Worthey94,Worthey97}. Over time, additional indices have been added to this list by other authors \citep[e.g.][]{Fanelli92,Kuntschner10,Du12,VidalGarcia17,Calabro21}.

While many modern high-multiplex spectral surveys now adopt a full-spectrum fitting approach, the Lick indices continue to encode much of the key information defining the stellar population contributions to the integrated light and are thus appropriate tools in which to compare the impact of assumed spectral library.

\begin{figure}
    \centering
    \includegraphics[width=0.48\textwidth]{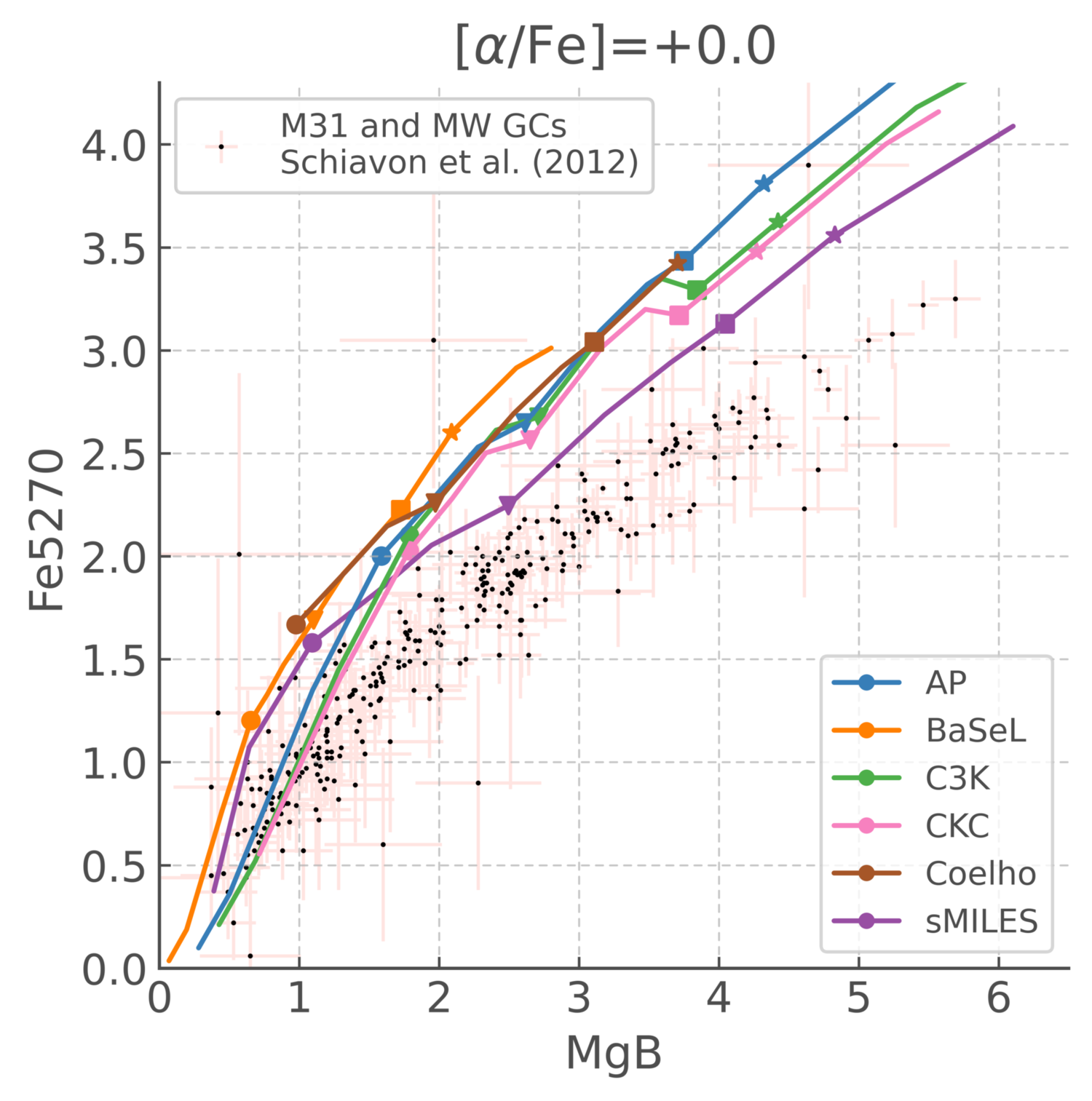}
    \caption{Evolution of Mg{\sc{ii}} and Fe\,5270 spectral indices as a function of metallicity for a simple stellar population with an age of 10 Gyr, with the different spectral libraries indicated by the different line colours. Black points show observational measurements by \protect\cite{2012AJ....143...14S} for globular clusters in the Milky Way and Messier 31, along with their uncertainties. The circle, triangle, square and star symbols on each line correspond to metallicities of 0.002, 0.006, 0.014 and 0.020 respectively.}
    \label{fig:MgFe}
\end{figure}

Fig~\ref{fig:MgFe} illustrates the evolution of the Fe5270 and Mg$_B$ spectral indices as a function of metallicity for a simple stellar population at an age of 10 Gyr. These indices primarily encode the strength of continuum line blanketing from iron-group elements, and the enhancement in $\alpha$-group elements respectively. Each of the spectral libraries is indicated on the figure by a different colour line and the circle, triangle, square and star-shaped symbols indicate metallicities of $Z=0.002$, $Z=0.006$, $Z=0.014$ and $Z=0.020$ respectively to aid comparison. While the model indices derived from all spectral libraries show a similar qualitative behaviour with increasing Z, there are some notable differences to be seen. The largest outliers are the BaSeL and sMILES libraries. This is not particularly surprising given their respective limitations; the undersampled low resolution spectrum of the BaSeL templates leads to an underestimate of the equivalent widths of absorption line features, while the normalisation of cool stars in the sMILES templates presented challenges due to their limited wavelength coverage, and these stars are important at $\sim10$\,Gyr. The remaining spectral template libraries all follow a broadly similar trend with metallicity. Nonetheless, the Coelho library produces index strengths which are consistent with a higher metallicity relative to the remaining libraries, and the AP library doesn't show the slight decrease in Fe5270 index strength which is present in the CKC14 and C3K libraries between metallicities of 0.01 and 0.014. 

\begin{figure*}
    \centering
    \includegraphics[width=0.76\textwidth]{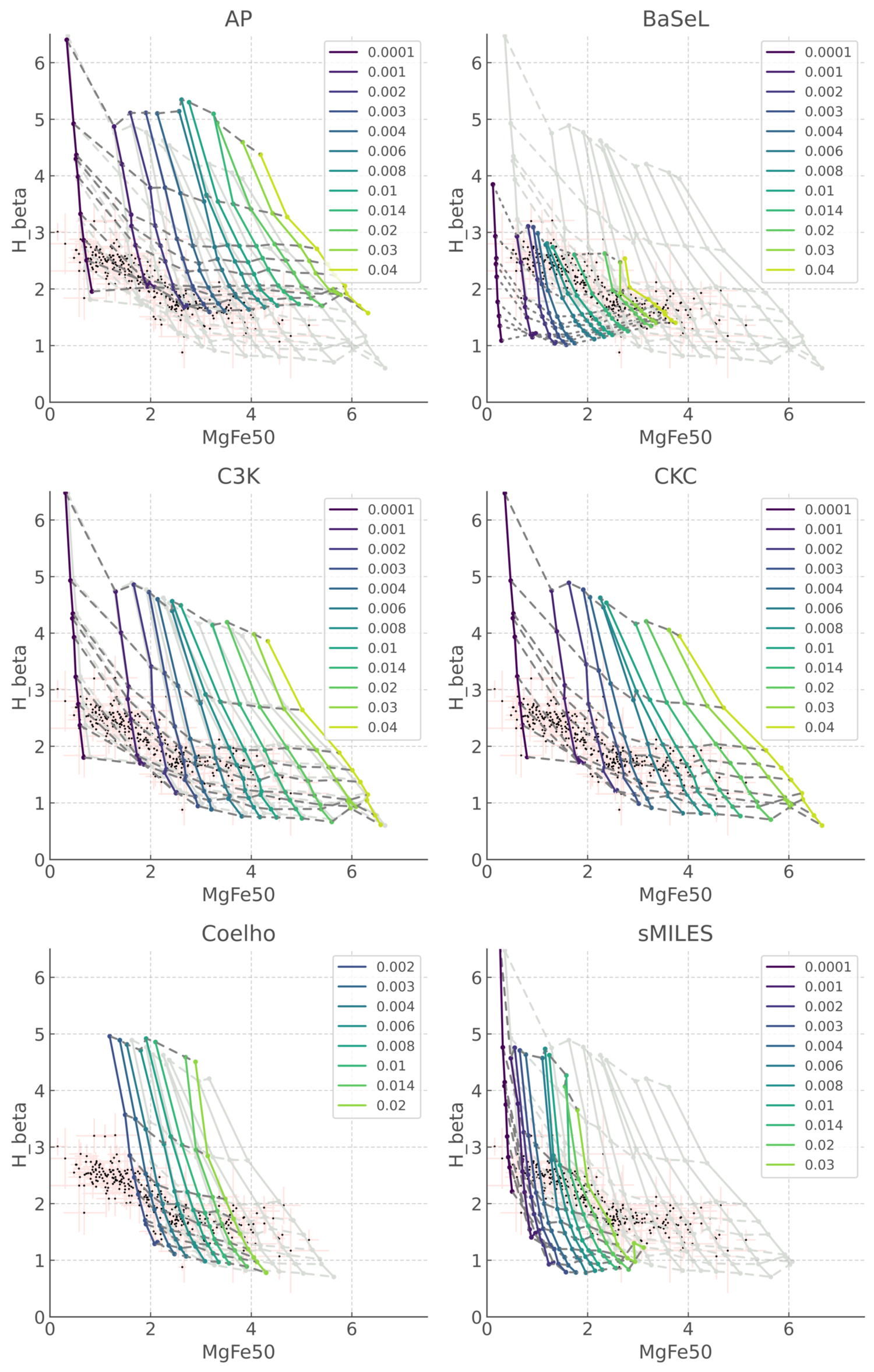}
    \caption{Strength of the H~$\beta$ and MgFe50 spectral indices as a function of age and metallicity for simple stellar population composite spectra derived using each of the stellar spectral libraries. Metallicities are represented by the different coloured lines, as indicated by the legend. Age points at different metallicities are connected by the dark grey dashed lines, corresponding to ages of 1, 2, 3.1, 4, 5, 8, 10, 12.6 and 15.8\,Gyr. After the upper left panel, each subsequent panel also shows the CKC14 library for the same age and metallicity range underplotted in light grey for comparison. Observational measurements of globular clusters in M31 and the Milky Way from \protect\citet{2012AJ....143...14S} are also plotted as black points, along with their uncertainties (pale red error bars).}
    \label{fig:hbetamgfe}
\end{figure*}

Fig~\ref{fig:hbetamgfe} illustrates the age and metallicity evolution of the strength of H~$\beta$ absorption and the MgFe50 index \citep[defined as MgFe50 = 0.5(0.69 Mg$_b$ + Fe5015) by][]{Kuntschner10}. This index was explicitly constructed to minimise the impact of uncertainty in $\alpha$-element enhancement on total metallicity estimates. The H~$\beta$ index measured the absorption line equivalent width of the Balmer series and hence is closely related to stellar population age.  The tracks indicate the time evolution of a simple stellar population from log(age/years)=9-10.2, while dashed lines connect tracks of different metallicity at constant age. The CKC14 library is shown as a reference in light grey on each of the subsequent panels. The \citet{2012AJ....143...14S} measurements of these indices in M31 and Milky Way globular clusters are plotted as black points.

Indices calculated using the C3K library span broadly the same region of parameter space as those using the CKC14 library, with a slight shift towards higher MgFe50 strengths at a given age (i.e. an observation placed on the diagram would have a lower implied metallicity than the CKC14 case), with changes in H~$\beta$ being typically less pronounced. The offsets are most extreme in the youngest, and highest metallicity, populations considered, reaching $|\Delta$MgFe50$|$=0.5\AA\ and $|\Delta$H$\beta|$=0.1\AA.

The sMILES and Coelho libraries both show a significant shift to lower values of the MgFe50 index, (i.e. an observation placed on the diagram would have a higher implied metallicity than the CKC14 case) but the H~$\beta$ index remains comparable. For a population with an age of 4\,Gyr and $Z=0.008$, the MgFe50 index would have values of 3.67, 2.85 and 1.65\,\AA\ in the CKC14, sMILES and Coelho libraries respectively, a difference of more than a factor of two between the lowest and highest values. Objects identified as Solar metallicity with CKC-derived indices would lie outside the modelled parameter range of both the Coelho and sMILES models and be interpreted as lying amongst the most metal rich of local systems.

Indices derived using the AP library give MgFe50 strengths which are comparable to the CKC14 and C3K libraries, but show an offset to higher H~$\beta$ line strengths (i.e. an observation placed on the diagram would have a higher implied age than the CKC14 case). Indeed, a number of the observed globular cluster population would have implied ages exceeding the age of the Universe. The Basel library-inferred indices cover a much smaller region of the parameter space. While the calculated indices broadly span the full range of observations (except for extreme examples in MgFe50), the implied ages show a strong trend with implied metallicity, spanning the full range of both quantities modelled. As was the case for the Coelho and sMILES libraries, the inferred metallicity of some systems would substantially exceed Solar. 

Additional visualisation of the differences between the spectral libraries in the photometric colours and selected spectral features are provided in the supplementary online material (labelled Appendix A) as a function of population age and metallicity. All differences are measured with reference to the results obtained using the CKC14 library, which was the fiducial spectral template library used in BPASS v2.2 \citep{2018MNRAS.479...75S}. 

\section{Effect of $\alpha$-enhancement on observables}
\label{sec:alpha}
Thus far, we have focused on results comparing the impact of spectral libraries with Solar-scaled compositions. As discussed earlier, $\alpha$-enhancement is widely observed and its underlying mechanism is an important process to consider, particularly in low-metallicity environments. Four of the six spectral libraries examined in this work, C3K, AP, sMILES and Coelho, provide $\alpha$-enhanced template spectra. While some libraries offer a number of different \afe\ options,  \afe=0.4 permits a direct comparison between all four of these libraries. In order to do so, the sMILES and AP $\alpha$-enhanced spectra were interpolated between adjacent \afe\ values.

In terms of photometric colours, changes are typically small (less than $\pm0.08$ mag for a given age and metallicity), except for in the bluest filters at late times, where the $\alpha$-enhanced models appear bluer. This is in agreement with previous studies of the impact of $\alpha$-enhancement on colours \citep{Vazdekis15,Byrne22}. In general, these differences are independent of library choice. Illustrations of the differences in colour predictions with $\alpha$-enhancement can be found in the supplementary online material (labelled as Appendix B). The mass-to-light ratio is mostly insensitive to changes in \afe, with variations on the $2-3$ per cent level. These differences are generally most pronounced at late ages and/or high metallicities. The 4000\AA\ break is unchanged at young ages, but becomes up to $\sim10$ per cent weaker for a given Z at late ages. This is consistent with old $\alpha$-enhanced populations appearing bluer, as seen in the colour information.

\begin{figure}
    \centering
    \includegraphics[width=0.49\textwidth]{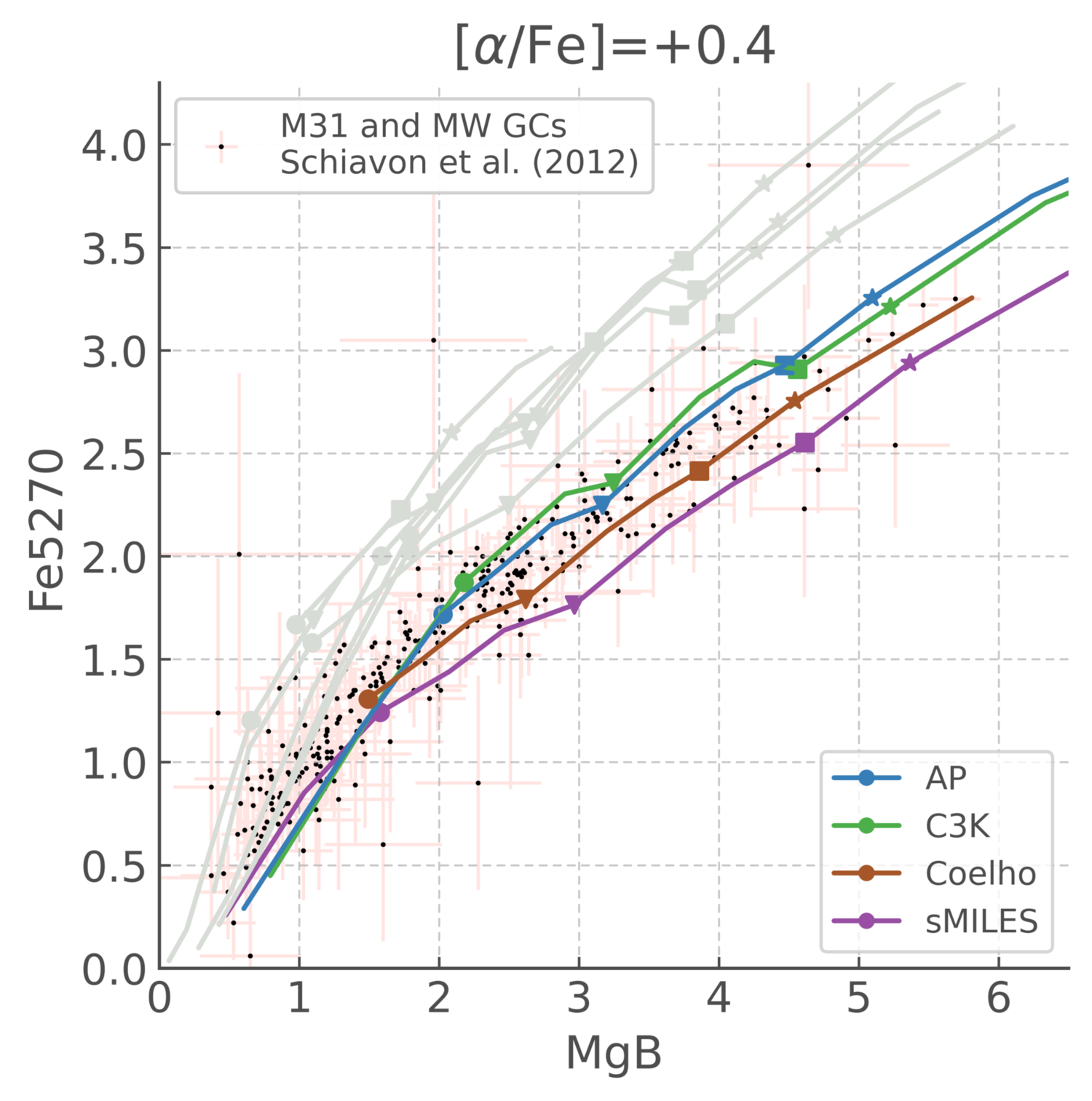}
    \caption{As Fig~\protect\ref{fig:MgFe}, but for the libraries which include $\alpha$-enhancement at \afe=+0.4. Once again, the circle, triangle, square and star symbols on each line correspond to metallicities of 0.002, 0.006, 0.014 and 0.020 respectively. The predicted indices from the \afe=0.0 models in Fig~\protect\ref{fig:MgFe} are shown in light grey for comparison.}
    \label{fig:MgFe_alpha}
\end{figure}

As might be expected, the impacts on probes of spectral line absorption are more pronounced. These manifest as changes in the behaviour of the Lick indices. Fig~\ref{fig:MgFe_alpha} illustrates the Fe\,5270--MgB Lick index relationship at 10\,Gyr, as in Fig~\ref{fig:MgFe}, but for \afe$=+0.4$. Once again, the \citet{2012AJ....143...14S} globular cluster data are also plotted. Through direct comparison with Fig~\ref{fig:MgFe}, it can be easily seen that these $\alpha$-enhanced models agree much better with the observations than the scaled-Solar composition models. This is not unexpected, since these old, metal-poor populations are expected to be enriched in $\alpha$-process elements. As before, indices derived using the sMILES library predicts a lower Fe5270 index strength at fixed MgB than the other models. The Coelho library spectral templates lead to a larger shift in the line indices with $\alpha$-enhancement than is seen in the CKC14 and C3K libraries. The size of the difference between the theoretical results using different libraries is typically smaller than or comparable to the uncertainties in the observational measurements.

\begin{figure*}
    \centering
    \includegraphics[width=0.85\textwidth]{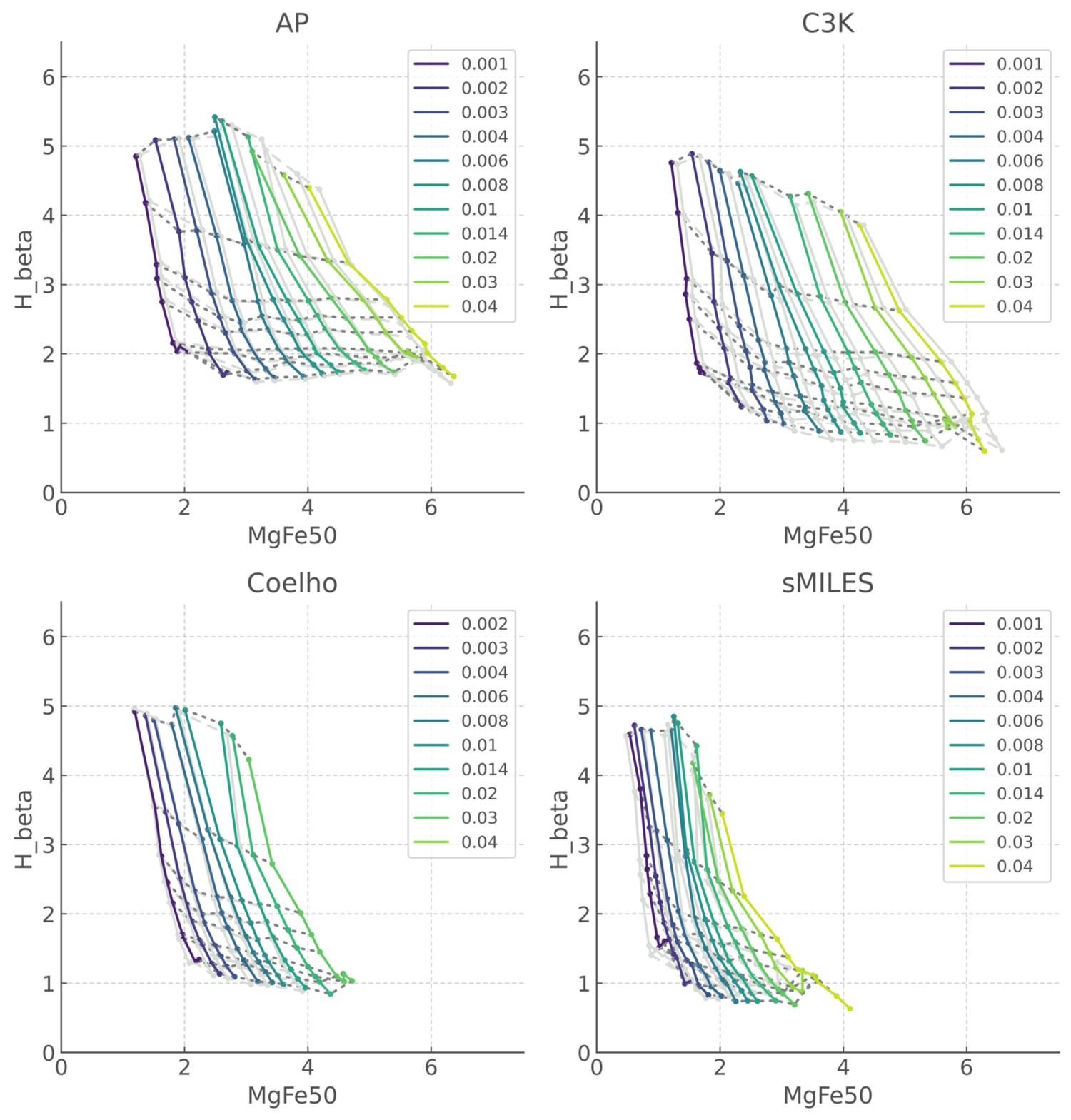}
    \caption{Similar to Fig~\protect\ref{fig:hbetamgfe} but for the $\alpha$-enhanced stellar spectral templates. Indices derived from the \afe=0 case in each stellar library are indicated by the light grey lines in the background.}
    \label{fig:indices_alpha}
\end{figure*}

Fig~\ref{fig:indices_alpha} illustrates the H~$\beta$ and MgFe50 indices as a function of age and metallicity for the $\alpha$-enhanced libraries. In each panel,the grey lines underplotted show the equivalent \afe=0 results from Fig~\ref{fig:MgFe} for the respective libraries. 

The H~$\beta$ line index contains no contribution from $\alpha$-elements or other strong metal line features and so should not be affected by $\alpha$-enhancement.
Since the MgFe50 index was explicitly constructed to minimise the impact of $\alpha$-enhancement on its constraint of metallicity from the line strength, little change is expected on either axis of this parameter space.  Indeed, Fig~\ref{fig:indices_alpha} demonstrates the \afe-independence of the Balmer line index. By contrast the MgFe50 index does show shifts with $\alpha$-enhancement, the magnitude of which depends on the stellar spectral template library adopted. The smallest impact is seen in the Coelho library-derived MgFe50 index in which the largest positive offset seen is $\Delta$MgFe50=4.6 per cent at $Z=0.002$ and age=16\,Gyr, while the largest negative offset seen is $\Delta$MgFe50=-4.5 per cent at $Z=0.01$ and age=1\,Gyr. The changes are less than 2.5 per cent for low metallicities and young ages ($Z\leq0.006$, age $\le$4\,Gyr) and high metallicities at all but the youngest ages considered ($Z\ge0.008$, age$\geq$1.6\,Gyr). The AP template library leads to offsets at young ages at all metallicities and at low metallicities across most ages. The largest offset is $|\Delta$MgFe50$|=7.4$ per cent at Z=0.020 and age=1\,Gyr. Only a small offset ($|\Delta$MgFe50$|<2.5$ per cent) is seen in models with $Z\ge0.04$ and an age $\ge$4\,Gyr. The sMILES library predicts indices which span a narrower range of MgFe50 than the other libraries, somewhat masking the impact of $\alpha$-enhancement. However the line strength is increased in the \afe=0.4 case at all ages considered in this figure, and particularly at low metallicity. The MgFe50 index is typically 10 to 15 per cent stronger in the $\alpha$-enhanced case for ages greater than 1\,Gyr, increasing with age. This increase in strength is largest at low metallicity ($Z\le0.004$).

The C3K template library generates indices which show the largest consistent offset in MgFe50 with $\alpha$-enhancement. At all ages and metallicities, models with \afe=0.4 show a $\Delta$MgFe50 towards lower values, with the offset ranging from 4.7 per cent at z=0.020 (age=10 Gyr) to 6.5 per cent at Z=0.002 (at the same age). The MgFe50 index combines a measure of the $\alpha$-sensitive magnesium absorption triplet at $\lambda\lambda 5167,5173,5184$ with a measure of the low ionization iron line blanketing of the continuum at $\lambda\approx5000$\AA. The index construction assumes that the strengthening of the former can be offset by the weakening of the latter for \afe$\ne$0. The C3K model grid was generated with updated absorption line lists and line strengths, based on the best available information in 2016. Compared to some other synthetic spectral templates, the iron absorption complexes have a bigger effect on modifying the detailed shape of the continuum. As a result, the iron component of the index may evolve more rapidly with composition than has been the case in older spectral libraries. Hence the fine balance between the two offsets which was defined using one spectral template set does not necessarily apply to all, and the scale and the direction of the offsets are determined by the line lists and atmosphere radiative transfer code adopted. These differences are illustrated graphically for H~$\beta$ and MgFe50 in Appendix B of the supplementary online material.

\section{Discussion}
\label{sec:discuss}
\subsection{Impact of Spectral Library Uncertainties on Observables}
\subsubsection{Photometric Properties}
The results presented here illustrate that the choice of stellar spectral library can have an impact on the predicted observational properties of a stellar population. Some observables such as the 4000\,\AA\ break and the mass-to-light ratio are relatively insensitive to the spectral templates adopted, their values and trends being dominated by the age and total metallicity of the stellar population. $M/L$ varies by no more than 7 per cent from the reference CKC14 spectral library (with differences typically less than 2 per cent). The 4000 \AA\ break, an important proxy for stellar population age, varies by no more than 4.7 per cent from the reference in the case of libraries with reliable results across the full age range. We note that this variation is consistent with the offset noted by \citet{Coelho20} between values derived using synthetic and empirical spectra.

Photometric colour predictions show good agreement both between spectral template libraries and with SDSS observations of galaxies dominated by either a young or an old stellar population. Photometric offsets between libraries are of order 0.08 mags in $u-g$ and 0.09 mags in $r-i$. As discussed in Section \ref{sec:alpha}, the variation in colour with $\alpha$-enhancement is similar in magnitude regardless of choice of spectral template library.

That this is a $<10$ per cent variation is to be expected, since the photometric colours of stellar populations are to first order determined by the thermal blackbody temperature of the most massive surviving stars at a given age, while the details of composition and resulting absorption lines provide a second order correction to the integrated flux on $\approx\,$2000\AA\ broad-band scales. Hence colours are relatively insensitive to both spectral template library and \afe\ selection, compared to the underlying population synthesis. All spectral libraries examined show similar trends with stellar metallicity.

However the offsets between libraries found here are a factor of 2-3 larger than those determined by \citet{Coelho20} for the difference between synthetic and empirical libraries implemented in the {\sc{galaxev}} spectral synthesis code. We note that a key difference between this study and the previous work, is the use of a full binary population synthesis, which precludes the implementation of smoothly varying, but overly simplistic, isochrones to describe the stellar population. As a result, the colour shows a stronger dependence on age and differences between libraries are amplified at ages associated with mass transfer or other binary interactions in massive stars (as can be seen, for example, in Fig. \ref{fig:colourcolour_sdss}).  

The offsets are also comparable to or slightly larger than the range in optical colour arising from binary fraction distribution uncertainty as evaluated by \citet{2020MNRAS.495.4605S}. The fact that neither spectral nor population uncertainties clearly dominate, emphasises the need to quantify these in both the population and spectral synthesis separately and using multiple techniques, before assuming any absolute `truth' in the relation between colour and population age.

The offsets between the colours predicted by different population and spectral synthesis models at a given age, are also somewhat larger than the photometric precision now attainable for galaxies at most redshifts at $z<7$, where a photometric uncertainty of a few percent or better is not unreasonable. Since this photometry is used for spectral energy distribution template fitting, the uncertainties on derived population parameters from such fits may be dominated by the differences in either spectral libraries used or population synthesis technique, rather than data limited. More work is clearly needed to understand and reduce the uncertainties on the theoretical side.

\subsubsection{Spectroscopic Properties}


We have limited the discussion of spectral properties in this work to well-defined spectral indices, rather than attempting full-spectrum fitting analysis. However these exemplify the trends and behaviour shown by the different libraries.

As shown in Section \ref{sec:output}, the model indices derived from all spectral
libraries show a similar qualitative behaviour. Outliers in the properties of line indices occur due to  under-sampled low resolution spectra
and  the normalisation of cool stars resulting from limited wavelength coverage. Quantifying these offsets presents a challenge, since the line indices are, by design, highly sensitive to age, metallicity and chemical enrichment.

However, while the trends in all libraries are similar, models derived using the Coelho library are consistent with the highest metallicity for a given line index strength. By contrast, those derived from the AP library would lead measured indices to be interpreted as the oldest stellar populations, in places exceeding the age of the Universe.

\citet{Coelho20} demonstrated how these line indices depend on the choice of empirical vs theoretical spectral libraries, in the simple single-star isochrone case. They found a typical offset between the libraries consistent with 0.07\,\AA\ equivalent width in H$\beta$, and -0.294\,\AA\ in the Fe5270 index (which probe age and metallicity respectively). This suggests that iron line blanketing may differ substantially between the spectral libraries considered, while the treatment of hydrogen recombination is, perhaps unsurprisingly, very similar.  They noted that the synthetic spectra typically yielded lower metallicity estimates for a given line index than their library of empirical spectra.

All the spectral libraries modelled in this work are theoretical. Compared to those considered by \citet{Coelho20}, they show comparable scatter in the Fe5270 index (see Fig. \ref{fig:MgFe}), but a larger scatter in the MgB index. This may reflect variation in the abundance pattern (rather than total metallicity) between the libraries considered. We note that all of the theoretical libraries would render low estimates of metallicity for a fixed Fe index by the criteria of \citet{Coelho20}, but caution that the empirical spectra are themselves, inevitably, calibrated against models to determine their metallicities, and the uncertainties inherent in this process must also be considered.

Any comparison of line index predictions against observational data must necessarily be conducted at the population statistics level, since multiple stellar populations which vary in age and metallicity substantially affect each measurement, and since the flux uncertainties on these narrow features are typically large. The uncertainty on  a single globular cluster data point, for example, will span several metallicity increments and age steps on our model grid.  However, the differences between theoretical models exceed even these large uncertainties, as Fig. \ref{fig:hbetamgfe} demonstrates. 

Of all the spectral libraries considered, the CKC14 and C3K libraries are the only ones to predict grids of spectral line indices which span the full range of observed data in the local Globular Cluster systems.

\subsection{Spectral Library Evaluation}

The C3K spectral library is the only library tested in this work which includes all three of: (i) wavelength coverage extending from the far-ultraviolet to the far-infrared, (ii) broad coverage of $T_{\mathrm{eff}}$-$\log(g)$ parameter space, and (iii) multiple variations of \afe.  As this work has demonstrated, the C3K library produces comparable predictions for stellar observables to other libraries tested, in regimes where the parameter coverage of all the libraries overlap. It predicts colours, $D_{4000}$ and other photometric properties which lie in the midst of other library predictions and which agree with observations. It produces spectral diagnostics which are broadly consistent with other libraries, but which demonstrate a slightly higher density of iron line blanketing in the stellar continuum. Again this is consistent with observational constraints. As a result, the robustness and parameter coverage gives it the best suitability to the diverse range of applications that \bpass\ is used for.

The AP spectral library also has significant parameter coverage, however issues such as the spectral break at $\sim9000$\,\AA\ in cool stars (which has no obvious physical origin), and the wavelength coverage (which stops short of the far-UV) limit the possibilities for exploring the uncertainties in young stellar populations. This shortcoming in UV wavelengths, particularly in the ionizing continuum of the far-UV, is common to all of the other spectral libraries. 

The BaSeL and CKC14 libraries present different issues when it comes to population and spectral synthesis.  The BaSeL library has a far lower spectral resolution than all other libraries considered here, and is also based on a since-superseded version of the stellar atmosphere modelling algorithms, which omit key improvements in our understanding. While the CKC14 library does not have these problems, it shares with the BaSeL grid an absence of any stellar spectra with non-Solar scaled abundance patterns. By contrast, all of the other libraries have one or more non-Solar $\alpha$-enhancements.

The absence of near-infrared wavelength coverage in the sMILES and Coelho libraries present substantial difficulties, in particular regarding the normalisation of cool star spectra, as shown in Section~\ref{sec:standards}. This is only one of several issues regarding spectral templates for low temperature stars, which are typically faint and extremely red, presenting empirical challenges, and also have turbulent, optically thick atmospheres, presenting theoretical challenges.


Given the large variance in spectral templates of the coolest stellar models, one potential source of offsets is the different low-temperature limits adopted by the individual spectral template libraries. As explained in section \ref{sec:popsynth}, in these cases, the \bpass\ spectral synthesis adopts the nearest available (i.e. typically hotter) stellar template at the same surface gravity. This particularly affects the sMILES and AP stellar libraries.

Arbitrarily restricting the libraries to only use stellar models at temperatures common to all libraries (i.e. T$_{\mathrm{eff}}\geq3500$\,K) should remove this effect, while only affecting relatively faint and low mass or extremely cool giant stars. In such a modified model, these will all be assigned the higher temperature, regardless of whether the correct temperature spectrum exists in the selected model library. We do this in a trial spectral synthesis grid, and find that it makes only a small difference to line indices and photometric colours, becoming more noticeable at high metallicity and at some ages, for example, the photometric difference between the AP and C3K libraries at Z=0.020 shows a noticeable shift towards bluer colours due to restricted parameter grid coverage at ages around $10^8$ years (when red supergiants can dominate the integrated light) and at ages above $10^{10}$ years (where cool dwarfs dominate). At Z=0.008, the differences only appear for red supergiants, while at Z=0.002 the models are indistinguishable. The differences between spectral libraries follow the same pattern as before, and remain larger than can be explained by this effect. This indicates that the differences between the libraries is more fundamental than a small artificial shift in the effective temperature assumed for the spectra of some cool stars.

\subsection{UV and Young Stellar Populations}

An area not explored in any of the results and comparisons presented here is the range of spectral features associated with very young stars. In the era of {\textit{JWST}}, it has become possible to probe to much higher redshifts than hitherto, observing these very young stellar populations. In these, the flux is dominated by hot stars, which emit a significant portion of their flux in the ultraviolet. Such populations will also typically be $\alpha$-enhanced. As very few of these spectral libraries extend into the UV (and only C3K both considers $\alpha$-enhancement and attempts to predict the ionizing spectrum) it was not possible to extend the detailed comparison carried out here down to shorter wavelengths. 

The prediction of the ionizing spectrum is essential since young stars are typically embedded in a relatively dense circumstellar medium, leading to the formation of H~{\sc{ii}} regions \citep{Xiao18}.  These regions, dominate the spectrum all the way into the optical, through a combination of nebular continuum and strong line emission. This nebular emission must be incorporated into any comparison between synthetic spectra and observations in the young stellar population regime. It is, on the other hand, largely irrelevant to older stellar populations, and thus has been neglected here.

We note that weaknesses remain even in the C3K spectral library version of \bpass\ (v2.3). In particular, the hottest stars require specialist atmosphere models which account for their radiative, diffuse and often-stripped envelopes. At present 
 \bpass\ adopts the Main Sequence/Giant Branch spectral libraries considered in this paper only for  $T_{\mathrm{eff}}\le25000$\,K, even though some of these libraries include spectra for stars hotter than this. The comparison made here uses the same subgrid of custom O-star and WR spectra for these hot stars. The inclusion of the hotter stars from each template library would go some way toward exploring uncertainties in young, hot stellar populations. New hot star spectra, analogous to the existing O star spectral library generated with {\sc{wmbasic}} \citep{PauldrachWMBASIC}, have recently been published \citep[e.g.][]{Hainich19}, while others may need to be generated. This would enable a direct comparison with the current hot star spectra used in \bpass.

\section{Prospects for the Future}
\label{sec:future}
\subsection{JWST}

The arrival of data from {\textit{JWST}} opens our eyes to the high-redshift Universe in unprecedented detail. Numerous high redshift galaxy candidates have already been identified in the Early Release Science data \citep[e.g.][]{JWSTAdams23,JWSTAtek23,JWSTCastellano22,JWSTDonnan23,JWSTFinkelstein22,JWSTFujimoto23,JWSTNaidu22,JWSTYan23} and in time, high resolution spectroscopic follow up will become available for many of these candidates. However, the spectroscopic subset will always be smaller than the photometric sample. As a result, the fitting of observed spectral energy distributions to template libraries will remain key to the interpretation of bulk galaxy properties including mass, age, star formation history and metallicity. Many SED fitting codes continue to use spectral templates tuned to reproduce galaxies in the local Universe accurately. Conditions in the distant Universe are known to differ substantially from those in the mature, nearby Universe \citep{Eldridge22Review}. As such, it is important that any SED fitting for these galaxies uses spectral templates which are appropriate for the expected properties (incorporation of $\alpha$-enhancement, young stellar populations, inclusion of binary stellar evolution) so that derived properties are reliable. The early, very young galaxies seen in {\textit{JWST}} and deep HST data will provide crucial constraints on the evolution of galaxies as a function of redshift, so it is important that the modelling uncertainties are accounted for, so that any conclusions drawn from such work is robust. This work provides a foundation for such modelling.

In addition to young stellar populations observed {\emph{in situ}} at high redshift, {\textit{JWST}} will also enable precision photometry and spectroscopy of middle-aged and old stellar populations at lower redshift. With sufficient data precision, the star formation histories of these galaxies can be explored, and galactic archaeology used to constrain the distant Universe and the evolution of galaxies in more detail. The near- to mid-infrared wavelength coverage of {\textit{JWST}} permits, for the first time, detailed studies of the rest-frame optical at intermediate redshifts, and the rest-frame infrared in local galaxies. Thus the wavelength coverage of stellar spectral libraries, which in turn determine the wavelength coverage of synthetic stellar population SEDs, is crucial to this work.

While SED fitting will be required for many distant galaxies, the sensitivity of {\textit{JWST}}'s spectrographs will also be key to the detailed interpretation of galaxy properties across a broad redshift range.  In particular, the metallicity, abundance indicator and age-sensitive lines targetted by the Lick and other spectral indices will be accessible for the first time beyond Cosmic Noon ($z>3$).  As we showed in section \ref{sec:indices}, the calibration of such indicators is subject to substantial variation depending on the adopted stellar library. This emphasises both the difficulty in obtaining an absolute calibration for these properties, and the importance of applying uniform methodologies when comparing different data sets and published results. While pixel-by-pixel full-spectrum fitting methods provide additional constraints, they will be sensitive to the same variation in line strengths between different template model libraries. These uncertainties are particularly pronounced when studying the oldest, and most metal rich, populations at any given redshift. 

\subsection{Future space telescopes}

Future space telescopes, such as the Nancy Grace Roman Space Telescope ({\emph{Roman}}\footnote{\url{https://roman.gsfc.nasa.gov/}}) and {\emph{Euclid}}\footnote{\url{https://www.cosmos.esa.int/web/euclid/}}, will be well equipped to examine distant galaxies in great quantities. {\emph{Euclid}} is predicted to produce images of around 1.5 billion galaxies and low resolution, near-infrared spectra of an estimated 25 million galaxies at redshifts of $1<z<3$, covering 15\,000\,deg$^2$ of sky to \citep{Sorce22EUCLID}. Meanwhile, predictions by \citet{Drakos22ROMAN} suggest that 1\,deg$^2$ ultra-deep field observations by {\emph{Roman}} (depth of $\sim30$\,mag) will detect more than 1 million galaxies photometrically, over 10\,000 of which will lie at redshifts consistent with the Epoch of Reionisation ($z>7$). Given the lack of high resolution spectroscopy in both of these instruments, the robustness of spectral energy distribution modelling, and the model-dependence of line strength indices, will continue to be an issue well into the next few decades. 

These instruments will significantly increase the number of observed galaxies at intermediate and high redshifts, providing constraints on galaxy evolution and permitted metallicity enrichment and star formation histories, but only if the bulk properties of the stellar populations can be determined with due precision, and devoid of systematic uncertainties. The results presented in this work suggest that we are not yet in a regime where this is the case. It is therefore of paramount importance that theoretical models continue to improve and sources of uncertainty are understood.

\subsection{Theoretical Advances}

This study highlights a number of areas in which improvements (to \bpass\ specifically and the field in general) can be made to further understand and reduce the uncertainties in the  spectra of synthetic stellar populations. This relies on improvements in the basic stellar template libraries employed for synthesis.

As discussed above, significant disagreement exists between different synthetic models of cool star spectra. These spectra are highly sensitive to the atomic and molecular line-list data included in the model, as there are very large molecular absorption bands present in the high opacity atmospheres of these stars. Further work, such as the consideration of non-LTE effects and calibration against nearby observed spectra should be done to bring these models into closer agreement. Here {\textit{JWST}} spectroscopy may help improve our understanding of low temperature stellar atmospheres through providing improved constraints in the near-infrared. Large differences in these spectra can play a role in the composite spectrum of a stellar population in regimes where cool stars dominate (i.e. red supergiants at intermediate ages and red dwarfs in old populations). 

The development of new tools for calculating the spectra of stars with atmospheres of arbitrary compositions \citep[e.g., {\sc{mps-atlas, }}][]{Witzke21} provides an opportunity to further examine the effects of non-Solar-scaled compositions, and  refine spectral synthesis models for stellar populations in the early Universe, in line with best estimates for galactic chemical evolution. Along with considerations of $\alpha$-enhancement in stellar spectra, some recent studies have also explored the impact of non-Solar-scaled compositions on stellar evolution models \citep[e.g.][]{Grasha21,Farrell22}. The \bpass\ stellar evolution code has recently been used to calculate small grids of $\alpha-$enhanced stellar evolution models, and analyses of these are under way. However the recalculation of the $\sim$250,000 binary interaction models which are combined in a BPASS population synthesis will take some time. Combining stellar evolution models and stellar atmosphere models which {\it both} use non-Solar-scaled compositions will allow for self-consistent modelling of stellar populations in low-metallicity environments.

As mentioned above, forthcoming observations by {\textit{JWST}} in particular will produce high-resolution rest-frame spectra of distant galaxies. This will eventually enable detailed statistical studies of galaxies at a broad range of redshifts, provided there are robust theoretical models with which to fit them. In particular, the high resolution in the redshifted rest-frame ultraviolet is likely to present a problem for most extant stellar population synthesis model suites. 

Presently, \bpass\ adopts a fixed 1\,\AA\ resolution for its output composite spectra, as a compromise to account for the varying resolutions of the numerous spectral libraries. Most modern theoretical spectral libraries (see e.g. Table~\ref{tab:libraries}) have a resolution much greater than 1\AA\ at UV and optical wavelengths. Indeed, this is one of the strengths of using theoretical spectral libraries over empirical ones. Future developments of \bpass\ will seek to increase the resolution of synthetic population spectra, so that more detailed spectral features can be studied in the output composite spectra. With the rise of full spectral fitting as a technique, it is important that the models have a similar or better resolution than expected from observations, to optimise the amount of science which can be extracted from the observations. 

\subsection{Future Approaches}

    In this paper we have concentrated on simple stellar populations in order to provide clear, direct comparisons between different libraries.   
    Another way in which the impact of theoretical uncertainties on interpretation can be demonstrated is through constructing the spectrum of an entire galaxy by combining multiple simple stellar populations. This must account for the age, metallicity and star formation history of the galaxy. This can be extracted, for example, from each pixel of a galaxy in observed data, or from the known properties of each particle in a cosmological simulation.
    
    By using a synthetic galaxy from a cosmological simulation, the results produced by using fitting to recover the inferred properties, based on each spectral template library can be evaluated. This approach will form the basis of a future publication from our team.

  Such an approach may also incorporate an analysis of the impact of non-stellar emission components.
  The models presented in this work have only considered the stellar component of an observed population. No contribution from the nebular gas which occurs ubiquitously in the neighbourhood of hot, young stars has been included. Similarly, no account has been taken of the presence in old stellar populations of diffuse ionized gas emission, powered by evolved stars. These introduce further uncertainties in the interpretation of observed data.



\section{Conclusions} 
\label{sec:conclusions}
The choice of stellar spectral library can lead to differences in the photometric and spectroscopic properties of a synthetic stellar population. This will directly impact determinations of derived ages, metallicities, and star formation histories of composite stellar populations from SED and photometric fitting of observations.

\begin{enumerate}

    \item Photometric properties such as colours and mass-to-light ratios, are perhaps the least sensitive observables to changes in the spectral library. Colours for simple stellar populations vary by up to 0.1\,mag at a given age. The scatter in colour predictions as a function of age and metallicity is consistent with the distribution of both dwarf star-forming galaxies and massive passive galaxies, as observed by SDSS. The magnitude of the uncertainty is comparable to the differences seen between empirical and synthetic spectra, and also to that produced by variations in the population synthesis model, showing that no one particular effect dominates the uncertainty. 
    
    \item Variations in the mass-to-light ratio are typically less than 2.5 per cent for young stellar populations and less than 7 per cent for older stellar populations. Excluding outliers, most libraries predict values of $D_{4000}$ which differ by less than 5 per cent.
    
    \item A lot of uncertainty remains in the spectral synthesis of cool stars, especially those with tenuous atmospheres, as different radiative transfer codes, opacity and line-list data can produce considerably different spectra of these stars. This can have a knock-on impact on the modelling of stellar populations in the age ranges where cool stars are the main contributor to flux, either as red supergiants at ages $\sim10^7$\,years or at late ages ($\sim10^{10}$ years) when low mass Main Sequence stars dominate.
    
    \item When looking at spectral line indices, the differences between stellar libraries can be quite considerable. Focusing on the H~$\beta$ and MgFe50 indices, which are sensitive primarily to age and metallicity respectively, large offsets are seen when comparing different libraries. The Coelho and sMILES libraries would imply a higher observed metallicity than the others, while the AP library would estimate older stellar populations than the other libraries. The low-resolution BaSeL library does cover the observed parameter space, but would estimate a much broader range of ages and metallicities than the other libraries. 

    \item The effect of $\alpha$-enhancement is noticeable in all libraries, with the behaviour generally being qualitatively similar. Colour index shifts with a change in \afe\ are very similar, regardless of library choice. Mass-to-light ratios and $D_{4000}$ show consistent changes also. The spectral line indices show a variation in behaviour, attributable to different treatment of iron line blanketing and atomic data.

\end{enumerate}

The C3K library is the only spectral template library used in this work which simultaneously provides the broad coverage of wavelengths, effective temperatures and \afe\ compositions necessary to begin to evaluate the young stellar populations that {\textit{JWST}} has observed at high redshift. However this study demonstrates that uncertainties remain in stellar templates in the optical, and these are likely still larger at longer and shorter wavelengths. Care must be taken when attempting an absolute rather than relative calibration of galaxy properties based on stellar population synthesis.

\section*{Acknowledgements}

We thank Jan Eldridge and the members of the \bpass\ team for providing helpful discussions and feedback while carrying out this research. The {\sc{hoki}} package \citep{HOKI} was used to extract and manipulate \bpass\ output data. We are grateful to Charlie Conroy and Ben Johnson for providing us with a copy of the C3K spectral library for use in this work. ERS and CMB acknowledge funding from the UK Science and Technology Facilities Council (STFC) through Consolidated Grant ST/T000406/1. 
This work made use of the University of Warwick Scientific Computing Research Technology Platform (SCRTP) and Astropy\footnote{\url{https://www.astropy.org/}}, a community-developed core Python package for Astronomy \citep{astropy:2013,astropy:2018}. 

\section*{Data Availability}

The output data from this work will be made available on the \bpass\ websites\footnote{\url{https://www.warwick.ac.uk/bpass}}\footnote{\url{https://bpass.auckland.ac.nz}}.


\bibliographystyle{mnras}
\bibliography{mybib} 


\bsp	
\label{lastpage}



\clearpage

\appendix
\setcounter{page}{0}
    \pagenumbering{roman}
    \setcounter{page}{1}

\section{Visualisation of the differences between libraries relative to the CKC14 reference library}
\label{Appendix:A}

\begin{figure*}
    \centering
    \includegraphics[width=0.99\textwidth]{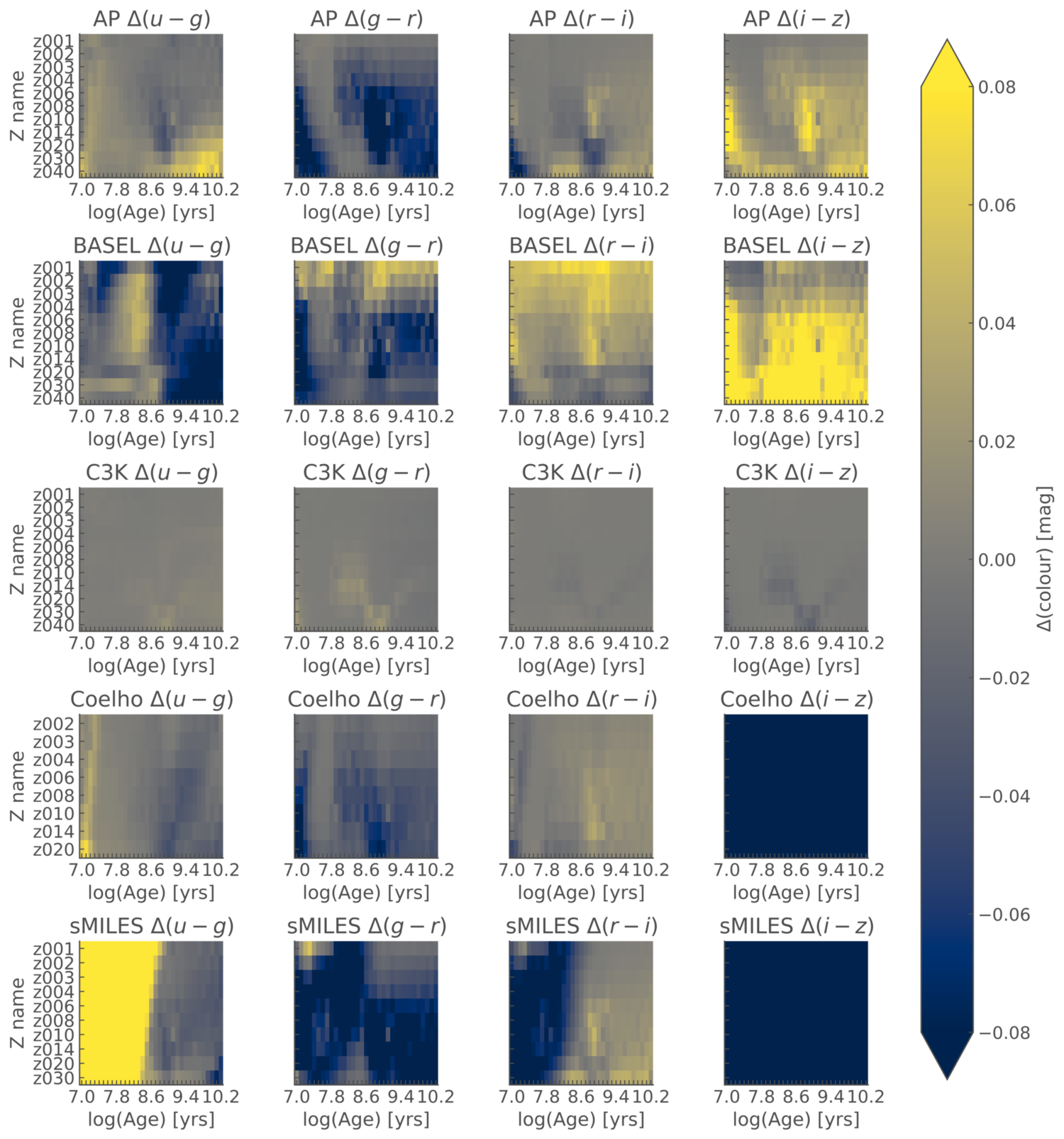}
    \caption{Differences between the predictions of the CKC14 spectral library and each other library  for colours $u-g$ (first column), $g-r$ (second column), $r-i$ (third column) and $i-z$ (fourth column). Data is plotted as a function of logarithmic age (x-axis) and metallicity (y-axis, labels correspond to $Z\times100$) for a simple stellar population synthesised using each of the spectral template libraries (one row per library, as indicated by the titles on each panel). Differences are stated relative to the CKC14 spectral library. The values of $\Delta(i-z)$ for the Coelho and sMILES libraries (fourth column, bottom two panels) show values far exceeding the colour scale due to the wavelength range of the$z$ filter exceeding the maximum wavelength of these spectra.}
    \label{fig:CKC_diff_Colours}
\end{figure*}

The figures in this appendix illustrate the differences between the spectral libraries for some of the key observable quantities present in the figures in the main body of the paper. Fig~\ref{fig:CKC_diff_Colours} illustrates the difference in colour-colour values from the CKC14 values for each spectral library and each adjacent pair of SDSS colour filters as a function of metallicity and logarithmic age bin in the \bpass\ output for ages between 16\,Myr and 16\,Gyr. The difference in colour, $\Delta\,C$, is determined as $C_{\mathrm{Library}}-C_{\mathrm{CKC}}$, so positive (negative) values indicate that a data point is redder (bluer) than the CKC grid at the corresponding age and metallicity. All panels use a colour scale extending to $\pm0.08$\,mag, which covers most of the variation seen. The sMILES library at young ages in $u-g$ and the sMILES and Coelho libraries in $i-z$ show much larger variations due to the absence of spectra hotter than 10000\,K and incomplete flux in the $z$ filter respectively, so these panels contain no meaningful data.

\begin{figure*}
    \centering
    \includegraphics[width=0.99\textwidth]{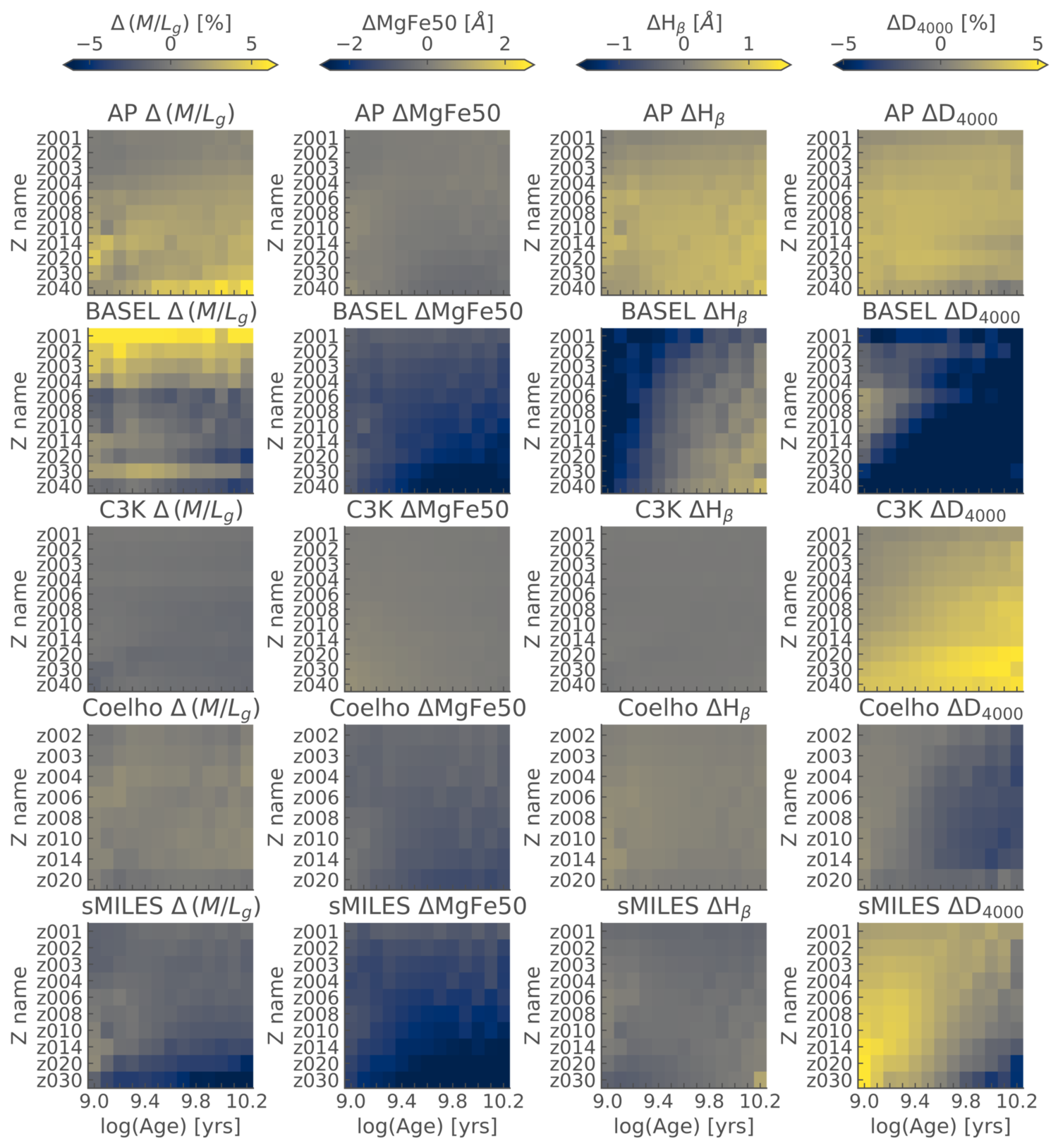}
    \caption{Differences between the predictions of the CKC14 spectral library and each other library  for $M/L_g$ (first column), the MgFe50 line index (second column), H~$\beta$ line index (third column) and $D_{4000}$ (fourth column). Differences are calculated as outlined in the text, and refer to a simple stellar population between the ages of 1 and 16\,Gyr. Each spectral library occupies a single row, as indicated by the titles on each panel.}
    \label{fig:CKC_diff_Indices}
\end{figure*}

Fig~\ref{fig:CKC_diff_Indices} illustrates the differences in the $g$-band mass-to-light ratio (first column), the strength of the 4000\AA\ break (fourth column), and the strength of the MgFe50 and H~$\beta$ line indices (second and third columns respectively) from 1\,Gyr to 16\,Gyr. The differences in $M/L_{g}$ and $D_{4000}$ are shown in terms of percentage change from the results predicted by the CKC14 spectra 
\begin{equation}
    \Delta X = 100 \frac{X_{\mathrm{Library}}-X_{\mathrm{CKC}}}{X_{\mathrm{CKC}}}.
\end{equation}

\noindent The differences in the line index measurements, $X_{\mathrm{Library}}-X_{\mathrm{CKC}}$, are given in units of \AA. The range covered by the colour scale is indicated at the top of each column. 

\section{Visualisation of the differences produced by $\alpha$-enhancement on a per library basis}
\label{Appendix:B}

\begin{figure*}
    \centering
    \includegraphics[width=0.99\textwidth]{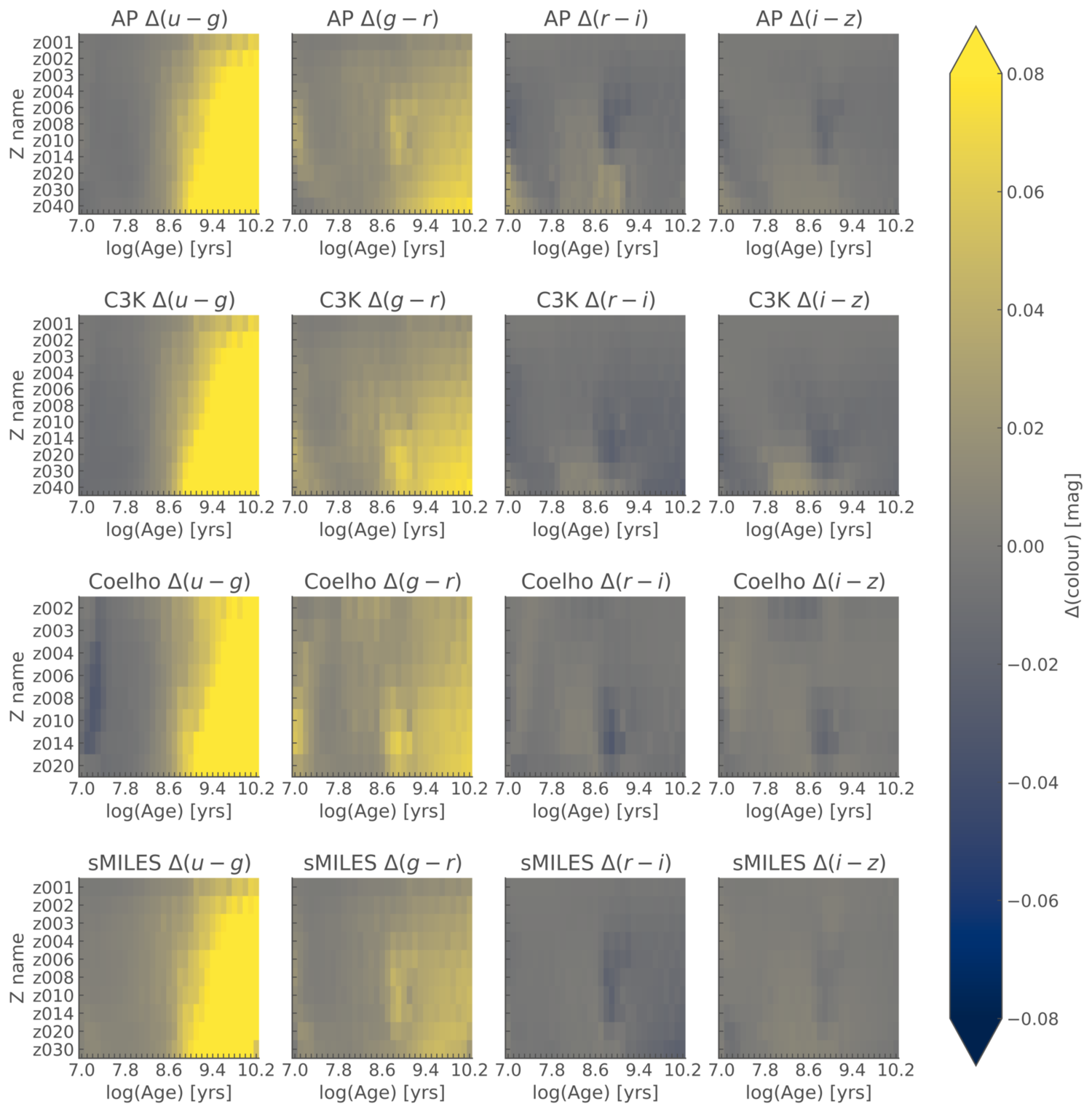}
    \caption{Differences between the predictions of \afe=0.0 and \afe=+0.4 spectra for colours $u-g$ (first column), $g-r$ (second column), $r-i$ (third column) and $i-z$ (fourth column) for each spectral template library (indicated by panel titles). Data is plotted as a function of logarithmic age (x-axis) and metallicity (y-axis, labels correspond to $Z\times100$) for a simple stellar population synthesised using each of the spectral template libraries (one row per library). Differences are stated relative to the \afe=0.0 case, as outlined in the text.}
    \label{fig:Alpha_diff_Colours}
\end{figure*}

The figures in this appendix illustrate the predicted differences between the \afe=0 and \afe=+0.4 stellar spectral templates within each spectral library. Note therefore that these difference values are not in reference to the CKC14 library as in Appendix~\ref{Appendix:A}, but, for example, the difference between the $\alpha$-enhanced AP predictions and the non-$\alpha$-enhanced AP predictions and so on. Fig~\ref{fig:Alpha_diff_Colours} shows the differences in SDSS colour-colour space $\Delta\,C_\alpha = C_{\mathrm{Library},{\alpha/\mathrm{Fe}=0.0}} - C_{\mathrm{Library},{\alpha/\mathrm{Fe}=0.4}}$. This means that positive (negative) values indicate that the $\alpha$-enhanced spectra are bluer (redder) than the equivalent Solar composition spectra.  All panels use a colour scale extending to $\pm0.08$\,mag, which covers most of the variation seen, and the known result that $\alpha$-enhanced spectra appear bluer at blue wavelengths and late times is readily apparent.

\begin{figure*}
    \centering
    \includegraphics[width=0.99\textwidth]{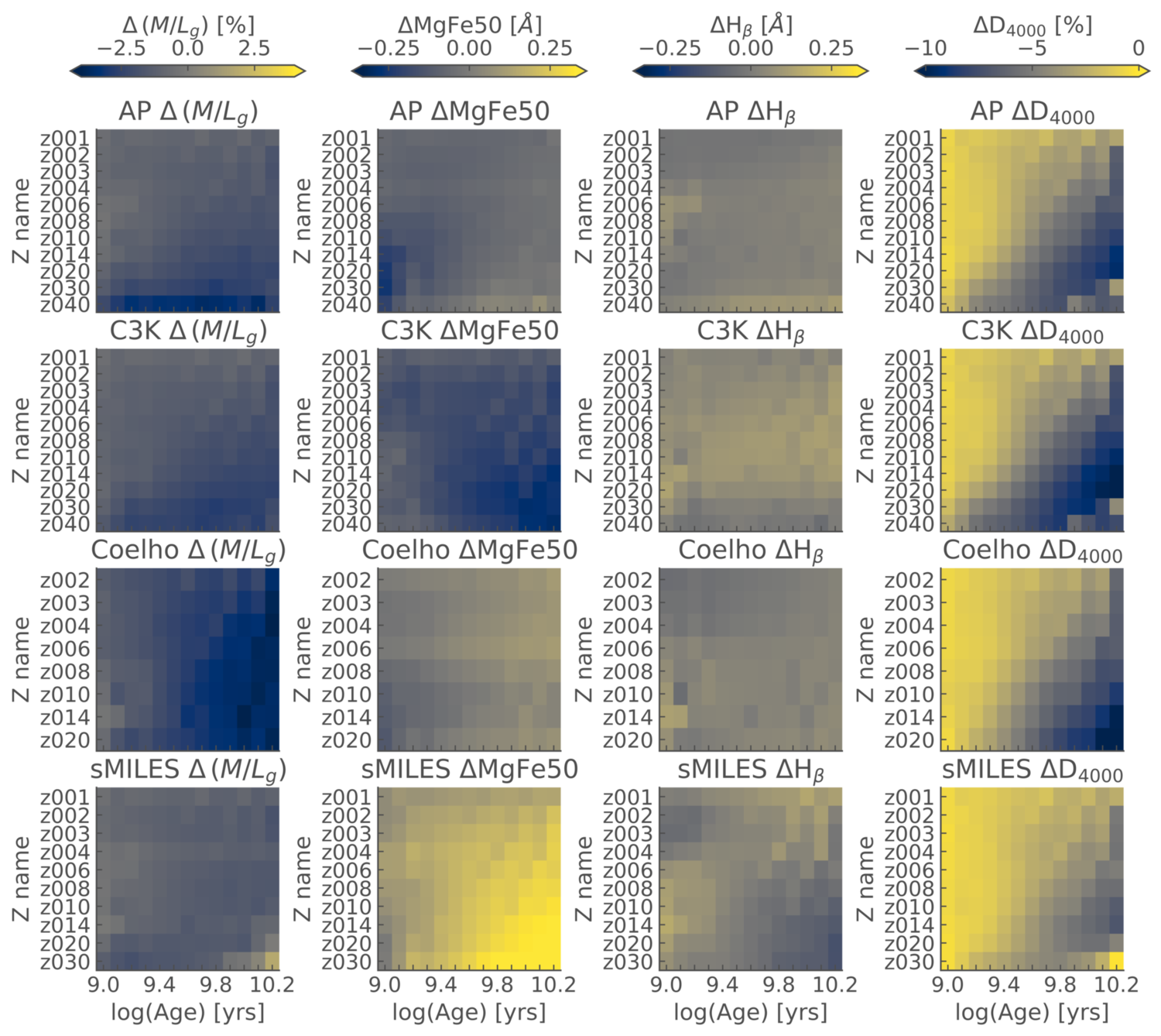}
    \caption{Differences between the predictions of \afe=0.0 and \afe=+0.4 spectra for $M/L_g$ (first column), the MgFe50 line index (second column), H~$\beta$ line index (third column) and $D_{4000}$ (fourth column). Differences are calculated as outlined in the text, and refer to a simple stellar population between the ages of 1 and 16\,Gyr. Each spectral library occupies a single row, as indicated by the titles on each panel.}
    \label{fig:Alpha_diff_Indices}
\end{figure*}

Fig~\ref{fig:Alpha_diff_Indices} shows the differences in mass-to-light ratio in the $g$ band, $M/L_g$ (first column) and $D_{4000}$ (fourth column) in terms of percentage difference:

\begin{equation}
    \Delta X = 100 \frac{X_{[\alpha/\mathrm{Fe}]=0.4}-X_{[\alpha/\mathrm{Fe}]=0.0}}{X_{[\alpha/\mathrm{Fe}]=0.0}}.
\end{equation}

\noindent The differences in the MgFe50 and H~$\beta$ line indices, $X_{\mathrm{Library, }\alpha/\mathrm{Fe}=0.4}-X_{\mathrm{Library, }\alpha/\mathrm{Fe}=0.0}$ (second and third columns), are given in units of \AA. The range covered by the colour scale is indicated at the top of each column. Note that the ranges spanned by these colour scales differ from those in Fig~\ref{fig:CKC_diff_Colours}.

\end{document}